# Focus3D: A Practical Method to Adaptively Focus ISAR Data and Provide 3-D Information for Automatic Target Recognition

John R. Bennett
Bennett Radar Software
1945 Briargate Place
Escondido, CA 92029
(858) 922-9732
mulebennett5@gmail.com

# ABSTRACT

To improve ATR identification of ships at sea requires an advanced ISAR processor – one that not only provides focused images but can also determine the pose of the ship. This tells us whether the image shows a profile (vertical plane) view, a plan (horizontal plane) view or some view in between. If the processor can provide this information, then the ATR processor can try to match the images with known vertical or horizontal features of ships and, in conjunction with estimated ship length, narrow the set of possible identifications. This paper extends the work of Melendez and Bennett [M-B, Ref. 1] by combining a focus algorithm with a method that models the angles of the ship relative to the radar. In M-B the algorithm was limited to a single angle and the plane of rotation was not determined. This assumption may be fine for a short time image where there is limited data available to determine the pose. However, the present paper models the ship rotation with two angles – aspect angle, representing rotation in the horizontal plane, and tilt angle, representing variations in the effective grazing angle to the ship. The mean values of these two angles are presumed to be known – aspect angle from a tracking process and the aircraft location, and tilt angle from the aircraft altitude and ship range. The time variations of the angles are estimated from a set of targets (bright points) on the ship that the ISAR processor uses to measure Range, Doppler (range-rate) and Acceleration (Doppler-rate). The 3-D method is based on only moments of the observed variables, and it does not attempt the difficult feat of trying to track every detected scatterer for longer times than the normal coherent integration time. The algorithm for focusing the ISAR imagery from these target reports is that of M-B. But the progress here is in the use of the focus information to model the aspect and tilt angles and then create scores for Profile, Plan and General 3-D poses. The pose scores are ratios of the variances of the XYZ coordinates of the targets in the natural (drydock) frame of reference and the variances of the noise assuming nominal uncertainties in Range, Doppler, and Acceleration. Focus3D is practical, requiring very modest computer resources and it can thus be implemented in most real-time ISAR processors. It also is self-evaluating in the sense that it monitors the validity of the 3-D solution to avoid times when the data is dominated by strong interference, or has multiple ships in view, or when the mean aspect and tilt angles are too small. The algorithm is illustrated and validated using a simulated data case, two classical cases from a shore test, and a large recent data set from a test radar.

# 1. Prologue: ISAR Focus via Two-dimensional Regression

The coordinates of an ISAR image are range and Doppler. Thus, the mean ranges and range-rates over the integration time of different parts of a ship are given by these coordinates; and the range acceleration is the focus function required to improve the contrast of the image and to accurately determine the shape of the ship and the location of features required for automatic target recognition. But determining the acceleration pattern over the image requires an adaptive algorithm. In Melendez and Bennett [1] this algorithm is based on the set of 'target reports', elementary bright points estimated from the data. Each target report consists of the estimated SNR, range, Doppler, Doppler-width, and acceleration of a scattering center for each time. The M-B algorithm for target detection uses the same image integration time as the main ISAR processor. Thus, the identification of the target reports could have been done using intermediate data products or the final complex image – the program allows either method. Either way, the focus function is determined by a data fit of the observed accelerations versus range, Doppler, and time.

The full M-B algorithm uses a robust algorithm to model the focus function. This algorithm uses 6 parameters to create a linear least squares (LLS) estimate of the acceleration versus range, Doppler, and time. This model is implemented via public domain matrix software after adding some small 'ghost targets' to remove the possibility of a zero determinant. However, this model can be understood as a variation of a much simpler LLS model that uses only two coordinates, range and Doppler. The logic here is that two of the terms in the full algorithm, the mean acceleration, and its time derivative, are redundant with other active algorithms in the ISAR processor, Adaptive Motion Compensation and Phase Gradient Autofocus. Also, the other time-dependent terms tend to be less important than the mean derivatives with respect to range and Doppler. In this reduced model there are only two parameters, Ar and Af – the derivatives of acceleration with respect to range and Doppler. The two-dimensional regression estimates of these are:

$$Ar = ( <a*r> * <f*f> - <a*f> * <r*f> ) / Det$$
$$Af = ( <a*f> * <r*r> - <a*r> * <r*f> ) / Det$$

Where the brackets indicate an average over the targets for a data frame and Det is the determinant of the regression matrix:

$$Det = <r*r> * <f*f> - <r*f>^2$$
$$\text{or}$$
$$Det = <r*r> * <f*f> * ( 1 - Crf^2 )$$

Here Crf is the Range-Doppler correlation coefficient, a quantity that is used in diagnostic plots for the program.

The program uses values of the covariance quantities in the expressions above that are scaled by the range variance. Thus, the Range-Doppler covariance CovRF is $<r*f>/<r*r>$ with similar scaling for the Doppler variance CovFF, the Range-Acceleration covariance CovRA, and the Doppler-Acceleration covariance CovFA. Note that for mathematical simplicity Focus3D program uses the m/s conversion of Doppler instead of Hertz. Similarly, on input the acceleration values are converted from Hz/s to m/s/s. Due to this scaling by the range variance, which allows the other four covariances to be expressed in terms of only the angles and their derivatives, there are only four basic quantities [CovRF, CovFF, CovRA, CovFA] necessary to complete the model.

This simple 2-parameter regression model naturally leads to the recognition that the key to 3-D ISAR is the modeling of these 4 covariance quantities. Furthermore, it is also useful to monitor the Range-Doppler correlation coefficient. Another useful derived quantity to monitor is simply called D = CovFF-CovRF^2, which is the intrinsic Doppler variance – the full Doppler variance minus the Doppler variance due to the angle of the ship in the range-Doppler plane.

Further analytical work has established that for 'perfect' data the two covariances involving the acceleration can be approximated by the primary covariances, CovRF and CovFF, and their time derivatives. 'Perfect' means that the same targets are identified in every frame of the dwell – this allows the interchange of the derivative and the expected value operators. Then the expressions of the time derivatives of CovRF and CovFF (CovRF_dot and CovFF_dot) can be rearranged to form the acceleration consistency relations:

$$CovRA = CovRF\_dot - CovFF + 2*CovRF^2$$

$$CovFA = CovFF\_dot/2 + CovFF*CovRF$$

For realistic data these expressions cannot be expected to be accurate since the number and position of the scatterers varies with time as targets fade in and out or interfere with each other. However, by comparing these synthetic estimates of the two acceleration covariances with either the data or the output versions of the lower moments we can form a convenient measure of the quality of the 3-D model. The parameter measuring this difference between the two measures versus time is called BadFit(t); it is used to avoid times in the data when the 3-D model is clearly not correct due to a failure of the rigid body assumption or a lack of detected targets. The program contains several alternate definitions for BadFit, some using the consistency measures and others using the differences between the data and output estimates.

Thus, the mathematical model involves two separate processes – (1) the estimation of the angles from CovRF and CovFF and (2) a validation stage where the output angle results of the first stage are used to test the consistency of the calculation of the acceleration covariances for both the data and the output angles.

## 2. Introduction

The Inverse-SAR Processor used here was initially developed in 1994. Since then, the algorithm has been applied to multiple ISAR systems. In the last 30 years I have collected or been sent many test results. I have permission to use these data sets for scientific work and algorithm development but not to publicize the systems or the data owners. Two of these data sets will be used here. First, a shore test from 1999 provided data from Marine Surveillance Radar #1 (MSR1). Second, a more modern test radar provided flight tests of Marine Surveillance Radar #2 (MSR2). All this data is from commercial ships – mostly container ships but a few tankers, bulk carriers and car carriers. Both data sets have a range resolution of 1 meter and a PRF of 1024 Hz, pre-summed to 512 Hz. All ISAR images and calculations presented here were done with my own Fortran and Matlab software from the raw data.

In the early implementations the algorithms in this processor used a least square fit to measurements of the motion of scattering centers detected on the ship. This method was described in Melendez and Bennett [1]. The method, called the Global Motion Model (GMM), estimates the rotation rate of the ship and its rate of change with time and then uses this information to estimate the true length of the ship.

The potential for three-dimensional Synthetic Aperture Radar was recognized in the fundamental paper on the polar format algorithm by Walker [2]. However, that paper concerned the deterministic processing for a static target but a highly maneuvering aircraft with a known path. This paper is concerned with an unknown motion of a ship at sea and thus the estimation of the motion is a critical question here.

Having had experience with many ISAR data sets, I became convinced that the quality of more recent data would allow a better motion model to be developed – that it is now possible to extend the Melendez and Bennett algorithm to the estimation of both the aspect rotation rate and the tilt rate of the ship. This would then allow us to estimate not just the length of the ship but also the height of the masts or the ship width, depending on the actual motion of the ship presented to the processor.

Much of the more recent work on 3-D ISAR [3-12] is concerned with the use of multiple apertures to support a direct estimation of mast height or ship width via interferometric methods. This method is a classical one that is well understood. However, only limited work has been done on applying it to ship imaging at realistic ranges and using realistic system designs. The purpose of the present effort is to do the best job possible using methods that can be used in existing maritime ISAR systems, which normally have a single aperture. This method could be extended to multiple apertures if future systems allow. But it is my belief that interferometric systems may have problems extending to ranges of hundreds of kilometers and that such a requirement is probably essential to most defense systems.

If we had a dual-antenna system with apertures 1 meter apart in the vertical, what would be the expected phase difference between them for a mast 10 meters high? I assume that the system would alternate transmission on the two apertures and receive on both – otherwise the interferometric baseline would be half the separation. The range difference between the two apertures is approximately $H*D/R$ where H is the height, D is the interferometric baseline and R is the range. At R = 100 km the difference is 0.1 mm – at X-band this gives a phase difference of 2.4 degrees. That is a small signal for a height difference of 10 meters. And the result gets worse linearly with range. Building and flying such as system would be a fair engineering and operational challenge for perhaps only a small gain.

But Focus3D may be able to achieve this accuracy without any further hardware modifications. The program works by using all the target data to estimate the angles and their time derivatives. Conceptually, the height offset is just the ratio of the radial velocity of the scatterer divided by the time rate of change of the tilt angle. Both Focus3D and a hypothetical interferometric method would probably only estimate height well for those frames with significant Doppler extent – for frames with a small Doppler spread the measured phase difference would apply to the sum of scatterers at multiple heights. And the Focus3D process with a single aperture works at any range where the system can produce adequate SNR.

It is possible that an interferometric ISAR with a horizontal separation would be more useful than one with a vertical separation. This would allow the angles to ship features to be estimated; and when combined with the range coordinate this can yield an estimate of the aspect angle, which can improve the critical ship length estimate. For slowly maneuvering ships the radar can estimate the aspect by tracking the ship. And for ships operating with the Automatic Information System (AIS) the transmitted course and speed can be used to calculate the aspect angle. Of course, if the ship transmits all the correct information, then ATR is not necessary since the ship ID is part of the AIS message. But an ISAR system needs to work for non-cooperating ships also since they may be hostile or involved in smuggling or illegal fishing. So, an interferometric aperture with a horizontal separation may substantially improve the system utility.

It is worth noting that the information from a vertical interferometer would need to be applied to each feature of the ship. However, a horizontal interferometer would mainly be used to improve the estimation of the aspect angle, a process that has two advantages. First, the angle difference would be larger since ships are normally longer than they are high. Second, the estimation of aspect angle only requires the derivative of the phase along the ship, a quantity that can estimated using all the scattering from the ship.

Interferometric ISAR is an intriguing concept, but it is difficult to implement in a real radar system and it is difficult to scale with range. The best solution for 3-D ISAR may be to use the Focus3D algorithm as a base but to enhance it with auxiliary information from any extra apertures that may be available.

## 3. Mathematical Basis of the Algorithm

A rigid body has three degrees of freedom, normally expressed as the yaw, pitch, and roll angles. However, if we assume that the bounces of the radar beam off the sea surface are not important and that the motion is small enough that it does not affect the scattering centers visible to the radar then the radar data can be modeled as only two angles which we call tilt and aspect. This geometry can be thought of as a fixed ship target with various scattering centers being imaged by an ISAR at some azimuth and elevation angles relative to the ship.

The Appendix illustrates the mathematical strategy behind the algorithm for estimating tilt and aspect. It starts with a few common assumptions like the range to the target is much larger than the target size. And it uses spherical coordinates to relate the target locations to the aspect and tilt angles and thus yield a description of the scatterer coordinates in a 'drydock' coordinate system - i.e., X, Y and Z are the locations of the scatterers in alongship, cross-ship and vertical. X, Y and Z are the desired answers to the 3-D shape of the ship. The conversion to spherical coordinates maps XYZ into the positions in range and in aspect and tilt angles. The first and second derivatives with respect to time of range yields the Dopplers and accelerations of the targets. The derivation further allows the covariances to be expressed solely in terms of the angles by dividing the raw covariances by the range variance of the data. This process provides a 3-dimensional matrix at each frame time that allows the conversion of range, Doppler, and acceleration to XYZ.

An additional assumption in the derivation is that cross products of XYZ can be ignored in favor of X*X, Y*Y and Z*Z. Also, the alongship coordinate is assumed to be larger than the beam or height values. However, the algorithm does estimate the ratios of <Y*Y>/<X*X> and <Z*Z>/<X*X>. Think of this as a narrow ship approximation with first-order corrections for the of width/length and height/length ratios.

In addition, the derivations show that the two angles can be determined from just two of the covariances, CovRF and CovFF. The two covariances using the acceleration, CovRA and CovFA, can be approximated by the two basic covariances and their derivatives in time. Therefore, these two covariances may be considered as auxiliary or validation data. Focus3D uses CovRA and CovFA as proxies for errors in the algorithm indicating inconsistency in the 3-D assumptions. For example, differences between the data and output versions of CovRA and CovFA are observed to be large when there is large interference from another radar or communications system or when there are multiple ships in the scene. This information is used to identify times when the 3-D model does not work – this guides the program to ignore these times when estimating the ship length or designating the frame types as Profile or Plan.

In illustrating the Focus3D results we will discuss three types of plots:

1. The Angle Figure, time plots of the tilt and aspect angles and their first two derivatives. For the simulated data we also plot the output against the known answer. This is the data used to evaluate the 3 X 3 motion matrix, a time dependent matrix that is used to convert the data on range, Doppler and acceleration to the 3-D drydock coordinates XYZ.

2. The Covariance Figure, time plots of CovRF, CovFF, D, CovRA and CovFA plus a plot of the Range-Doppler correlation coefficient. This presents the evidence that the angle output of the Angle Figure is consistent with the observed covariances.

3. The Acceleration Consistency Figure, time plots of CovRA and CovFA versus their approximations in terms of the lower moments – with separate plots for the raw data and for the algorithm angle output.

## 4. Testing – Ideal Case

The Focus3D code contains two different methods of simulation for testing. First, it allows the generation of perfect input data – the exact values of range, Doppler and acceleration calculated from the mathematics for a model ship. This type of simulation is not illustrated here but it is very valuable for code testing. The second type of simulation, where the simulated targets are used to generate ISAR signal history to drive the ISAR processor is shown here. In this simulation not all targets are detected because there is interference between targets that are nearby in range or Doppler. The raw data also has variation in radar cross section over the targets and added white noise and thus yields a large range of SNR values.

Figure 1 shows the angles estimated from the program for a case that has a steady turning rate in aspect and two oscillatory components, a 12-second oscillation in aspect and a 10-second oscillation in tilt angle. The mean value of aspect is 45 degrees, and the mean value of tilt is 30 degrees. The 3-D answers are plotted in blue, and the perfect solutions are plotted in red.

Figure 2 shows the results for the moments used in the estimation of the aspect and tilt angles plus the range-Doppler correlation coefficient. The covariances from the data agree with those from the output angles.

Figure 3 summarizes the results for the auxiliary moments that are not used in the algorithm for estimating the aspect and tilt angles. There are four sub-plots that compare the Range-Acceleration and Doppler-Acceleration covariances with the perfect solution. The plots at the top are for CovRA and the plots on the bottom are for CovFA. The plots on the left are for the input data and the plots on the right are for the output data. This verifies that the covariances using the acceleration data can be approximated by the lower order covariances and their time derivatives.

## 5. Testing – Shore Test Data

For this paper we will illustrate the algorithm with two interesting cases collected during shore tests with MSR1 and multiple airborne cases with MSR2.

MSR1 Cases S00754 and S00759 are ships of opportunity collected while testing the radar from a shore site. Case S00754 is a large ship (about 75 meters long) with little superstructure – possibly a tanker or cargo ship. We refer to Case S00754 as **Large Ship Turning**. From photos taken from the shore we believe that Case S00759 is a bulk carrier of the same class as the SANKO MAJESTY (Figure 4). The shore test photos are not high quality but show the name of the shipping line, **SANKO**, and the 4 large cranes, the deckhouse, and the bow mast with weather and communication equipment.

The motion for **Large Ship Turning** was dominated by a steady turning rate of about 0.7 degree/sec in azimuth. **SANKO** had only a small turning motion but a very regular tilt component.

For **Large Ship Turning** Figures 5-7 repeat the three plots used for the simulated case but with no perfect data. We do not have measurements of the actual rotation rate of the ship, but this result is consistent with a qualitative visual analysis of the ISAR movie. Also, the length of the ship is relatively constant in time, indicating that the cosine-aspect factor is correct versus time.

For **SANKO** Figures 8-10 repeat the three plots used for the simulated case but with no perfect data. Again, this is a ship of opportunity that appeared off the test site, so we do not have direct measurements of the tilt angle to the radar versus time. However, a manual replay of the movie in single frame mode shows that the masts flip quasi-periodically from positive to negative Doppler in sync with the values of the tilt rate from Figure 8.

So, the three examples of a simulated ship, a constantly turning ship, and a ship with periodic tilting give some confidence that the algorithm is on the right track and that it deserves a more detailed validation process. As we will show, the 43-case Length Exercise Data Set confirms that the algorithm's answers are reasonable. Also, this set of cases is large enough to allow us to observe the rare events where the model fails due to external problems, thus allowing us to develop bullet-proofing methods to avoid the worst effects of these problems.

## 6. Profile and Plan Images

Section A.2 of the Appendix derives the fundamental 3 X 3 motion matrix that converts the input data on Range, Doppler, and Acceleration (RDA) to the drydock coordinates XYZ. This matrix is a function of time and is composed only of functions of the aspect and tilt angles and their first and second derivatives. In using this matrix Focus3D carefully monitors the matrix condition number to avoid times when it is close to singular. For other times the inverse of the motion matrix is multiplied by RDA to yield XYZ.

The RMS noise for XYZ is estimated by applying the inverse of the motion matrix to the nominal estimates of the error bars on RDA. For example, the error bar on range is a fraction of the range impulse response. Given the image integration time T, the error bar on Doppler is a fraction of $1/T$ and the error bar on acceleration is proportional to $1/T^2$. The program then computes scores for Profile and Plan image frames as Variance(Z)/NZ and Variance(Y)/NY where NY and NZ are the estimated noise variances for Y and Z.

Any frame that exceeds a threshold value is designated as a Profile or Plan frame; and the code also identifies general 3-D frames that have high scores for both. A fourth type of frame – one that is of limited usefulness for an ATR algorithm – is called String-of-Pearls. In this type all the targets are on a line in range Doppler space and there is no Doppler information to exploit. However, a final determination of this designation is only done if the score is the highest of the scores for all frame types – it does not make sense to assert that a frame could be both Profile and Plan.

All the designated frames are available for processing by an ATR algorithm. However, the program also produces a composite image for Profile and Plan. The composite images are formed from the sum of all the frames of each frame type after an auto-registration process that uses a 'rubber sheet' algorithm. The composite images are formed from the original range-Doppler images that are the output of the ISAR processor. But the coordinates of the composite images are range and cross-range in meters since the conversion from Doppler to meters uses the tilt rate for Profile frames and the aspect rate for Plan frames.

Figure 11 displays the composite image for the simulated case. This composite used 14 Profile and 9 Plan frames; no frames were designated 3-D. These results are very close to the shape of the simulated ship.

Figure 12 displays the composite images for **Large Ship Turning** and **SANKO**. Although we do not know the exact answer for these ships, the ISAR movies clearly are consistent with a steady rotation for **Large Ship Turning** and a variable tilt motion for **SANKO** with the mast signatures oscillating from negative to positive Doppler. Focus3D has a method that detects the sign flips when the signature goes through times when there is zero range-Doppler correlation. Therefore, the composite process inverts either the positive or negative values of the angle rates to align the images. The program detected zero Profile frames and 19 Plan frames and zero Profile frames for **Large Ship Turning** and 30 Profile frames and zero Plan frames for **SANKO.** These results are consistent with what an observer of the ISAR movies would probably decide.

## 7. MSR2 Data

The Focus3D algorithm clearly needed real life testing for a large sample of ships and a variety of sea states. Therefore, I acquired a large set of data cases from MSR2, collected in October 2019 and June 2020. The 2019 data used a forward look, and the 2020 data had a left look. After removing cases where the radar beam missed the ship or imaged other features like bridges there were about 200 cases remaining. Those with at least a signal-to-clutter ratio (SCR) of 6 dB were selected and these cases were culled further based on common sense considerations … low dwell time, small targets, Over-the-Horizon, high radar interference, weird ship signatures, lack of Ship ID from the system, signal fading, etc. After the culling there were 23 cases from the October 2019 tests and 20 from the June 2020 tests, which were designated the Length Exercise Data Set. Almost all these cases had auxiliary recordings from the ships' Automatic Information System (AIS) transmissions. This data allowed us to test the length accuracy of the algorithm independent of the accuracy of the radar tracker. The Length Exercise Data Set consists of commercial ships – mostly container ships but some tankers, RORO (roll-on, roll-off), a small cruise ship, and other types.

We will first use the MSR2 data to illustrate the BadFit algorithm for detecting times when the 3-D logic does not work well due to extraneous ships in the scene or interference. Then we will compare the 3-D result with the auxiliary data estimate of the mean rotation rates of the ships. Then we will evaluate the accuracy of the Focus3D length estimator and discuss the variations in the minimum and maximum range estimates. Finally, we will review selected composite images for profile and plan views.

## 8. Practical Enhancements – The BadFit Parameter

A key word in the title of this paper is 'practical'. A practical algorithm should automatically detect and react to the unavoidable effects that occur in data collection at sea. As discussed in the prologue, this is done via the BadFit parameter. The Length Exercise Data Set has several clear signatures that illustrate this algorithm.

For example, the radar data could have confusor ships in addition to the desired one. Figure 13 shows an example of this – there is a bogey ship that initially occurs near the same range and Doppler of the target ship. The bogey moves through range rapidly and eventually leaves at the far range. This rapid motion at the same nominal Doppler indicates that the bogey is at a Doppler alias velocity of the target ship.

In addition, we have identified two distinct signatures of interference from another radar or communications system. These features are completely misfocused in range – probably because they have a narrow bandwidth and/or are not matched to the chirp of the radar.

The first interference type, illustrated in Figure 14, is a pattern that is visible at every time but has a relatively narrow Doppler extent. This signature migrates through Doppler over time due to a combination of its intrinsic characteristics and the motion compensation applied to focus the ship. When the band is near or over the ship the target detector has trouble detecting the parts of the ship.

The second interference type, illustrated in Figure 15, is a pattern that occurs briefly in time but has a relatively wide Doppler extent. This signature essentially increases the noise/clutter level, making it difficult to identify bright scatterers on the target ship.

The Focus3D algorithm handles the bogey problem and the two interference problems with a common algorithm that measures the goodness of fit of the range-Doppler moments from data and from the output angles. This algorithm, called BadFit, was developed in a process of experimentation with alternate measures. The key method is largely based on the fit to the moments that involve the accelerations; but it also uses the fit to the Doppler moments. When the BadFit parameter is large the algorithm responds in three ways. First, it increases the implied noise in determining the Profile and Plan frames from the inversion of the 3-D motion model. Second, it ignores these times in the length estimation process. Finally, the program computes global performance scores in two ways, either using all the image frames or ignoring the BadFit times.

Figure 16 shows time plots of the BadFit parameters for the above three cases. The top sub-figure is for the case (Figure 13) with the bogey in the scene. This extra feature is clearly aliased in Doppler – it has an actual range velocity that corresponds to the wrapped Doppler close to our ship but with an actual Doppler off by some multiple of the PRF. This causes a non-rigid body signature that throws off the 3-D logic. Finally, the bogey either moves out of the image or it separates enough from the main ship that it is rejected by the program's outlier rejection process. The middle sub-figure is for the case (Figure 14) with a sharp line is Doppler that migrates through the scene over time. This does not give the program any problems until the interference line passes through the ship. At this time the BadFit parameter peaks and eliminates the bad frame from further analysis. The last sub-figure of the figure is the BadFit for the case (Figure 15) with broadband Doppler noise that varies episodically in time. The bad regions of time occur at many different times and when they peak the BadFit parameter adjusts to down-weight those times.

## 9. Mean Aspect Rotation Rates

The Focus3D algorithm assumes that the low frequency response of the ship is dominated by a nearly steady rotation in azimuth. Figure 17 compares the 3-D estimate of the mean rotation rate to that calculated from the system auxiliary data. As could be expected, the large values of rotation rate agree well – most of these large values are for cases with short range and larger tangential aircraft velocity. But for longer range or slower aircraft velocity there are larger differences. This difference is probably a reflection of the difficulty of estimating the ships' tracks at longer range.

## 10. Ship Length Estimation

Focus3D contains a ship length estimator that has the following characteristics:

1. The algorithm is based on the Thin Ship Approximation where the estimator is the Range Extent (maximum range minus minimum range) divided by the product of the cosines of aspect and tilt angle.

2. The algorithm has a method for reducing multipath effects – mostly ghost scatterers at longer range due to internal reflections among the cranes on a bulk carrier or among containers on a large container ship.

3. It has a correction for the width of the ship that depends on the aspect angle – basically to account for the fact that either the bow or stern is wide enough to cause an increase in the apparent length.

4. The width correction is significant for larger aspect angles since the first or last scatterer detected may be one of the corners of the stern. This correction requires an estimate of the stern width. To estimate the width, we tried to use the 3-D solution but sometimes the radar can only see one side of the ship. It was determined that it is better to use the following Naval Architect's Rule-of-Thumb to estimate the width from the first guess of length.

   [https://en.wikipedia.org/wiki/Beam_(nautical)]

5. The algorithm uses a double median measure for the LOA estimates versus time to eliminate stray large values due to extraneous signals like uncompensated multipath or reflections from other ships.

Figure 18 is a plot of the length overall (LOA) errors for the 43 Length Exercise cases versus the estimated Signal-to-Clutter Ratio. There is essentially no trend with SCR, although one would expect that higher SCR would lead to lower error. This means that we could have probably set a lower SCR threshold than 6 dB and thus get more good cases. The RMS LOA error is about 8% (19

m) and there is about a 3% negative bias. The negative bias is understandable because sometimes the ends of the ship do not always have large scatterers. Also, the far end of the ship might be affected by shadowing by the superstructure. Positive LOA errors could happen in cases where there is multipath scattering, which causes scattering beyond the far end. However, the code has an algorithm to defend against multipath. And in any case, we have found multipath to be rare, usually occurring when there is a large structure like a deckhouse that acts as a mirror and large scattering centers inside the body of the ship (e.g., the tall cranes on some bulk carriers).

Figure 19 plots the values of the standard deviation of the minimum versus the maximum range over time. The main result is that the minimum range target has much less time variability than the maximum range. This is probably due to multipath scattering and superstructure shadowing of the far end of the ship compared to the ease of detecting the first target against a background of open sea. This has implications for any ATR algorithm – it may be more effective to measure features relative to the minimum range than the far range or the mean range.

## 11. MSR2: Selected Composite Images

Figures 20-22 show photos of three ships along with the output composite images from Focus3D.

The first ship is the **MONTE ROSA,** a 272-meter container ship. This ship was imaged at a range of 101 km, an aspect angle of 21 degrees, and a grazing angle of 0.5 degree; it has a mean SCR of 12.5 dB. Focus3D detected only Plan view image frames and thus did not output a composite Profile image.

The second ship is the **SOLVIKEN,** a 249-meter tanker ship. This ship was imaged at a range of 93 km, an aspect angle of 9 degrees, and a grazing angle of 0.4 degree; it has a mean SCR of 7.2 dB. Focus3D was able to detect both Profile and Plan frames. The two composite frames appear to agree with the deck and vertical features visible in the photo.

The third ship is the **TENO,** a 299-meter container ship. This ship was imaged at a range of 175 km, an aspect angle of 45 degrees, and a grazing angle of 0.8 degree; it has a mean SCR of 8.6 dB. This is the case mentioned above where the ship was imaged during a sharp turn. Thus, although the far range would tend to reduce the rotation rate due to the aircraft motion, the actual rotation is quite large, causing the ISAR imagery to be dominated by Plan views. Thus, the program did not identify any Profile views.

I have done a detailed review of all 43 cases for 32 ships plus I have analyzed several other data cases from MSR2 radar and other systems. This review showed that we can expect a wide range of results. Some cases provide many selected frames and other cases provide few. Normally the results make sense compared to a manual selection process but occasionally there is some crosstalk between the frame types, with both composite frames looking similar. Note that I would expect an ATR algorithm to work directly with the individual frames instead of the composite frames. The ATR models for each frame type may have overlap so this may not be a serious issue.

## 12. Full circle: Return to Discussion of the Focus Parameters

In the Prologue we discussed the relation of the autofocus parameters used in the ISAR processor to the moments of the target statistics. The ISAR processor used here is somewhat more complex, modeling the time dependence of the fit in addition to the dependence of acceleration with range and Doppler – but this is a detail since we can understand the process well enough with the two-parameter regression model. This reduced model of the autofocus acceleration field uses a two-dimensional regression fit of the target accelerations to the ISAR image coordinates, range and Doppler. The derivatives of the acceleration in range and Doppler are Ar and Af. This derivation naturally led to the identification of the important covariances that motivate the 3-D model and which we have illustrated in detail for simulated and real data.

We can close the logical circle by comparing the curves for Ar and Af from the original ISAR processor with the versions computed by Focus3D. Figure 23 presents these results for **SANKO**. The upper-left section of each figure is the plot of the range-Doppler correlation coefficient. The upper-right section is the plot of the intrinsic Doppler width. The lower-left panel is the results for Ar – the original data from the ISAR processor, the program result for the input data, and the program result for the output (i.e., the estimate of Ar from only the angle results). The bottom-right panel is the similar result for Af.

The two lower plots cases show nearly perfect agreement between the GMM values from the ISAR processor and the Focus3D calculation from the input data (the blue and green curves). However, for both Ar and Af the calculation from the angle outputs tends to be spiky. The basic pattern of the curves is the same as for the ISAR processor and the Data calculation but there is a tendency to overshoot at the times when there is a sign flip. This is interesting but not very concerning. This happens when the range-Doppler correlation peaks and the intrinsic Doppler width is lowest. And according to the equations in the Prologue the determinant goes to zero when the correlation goes to 1. The blue and green curves agree since they are both are calculated with a defensive algorithm that nulls out the effect of the small determinant. In any case, these are the times that we call String-of-Pearl frames – where all targets are on a line in the range-Doppler plane and the movie is in transition between plus and minus Doppler. These times are not useful for ATR purposes since all the Doppler information is collapsed to a line … there is very little image structure to exploit. Mathematically, the Out patterns tend to exhibit this behavior because by construction these curves are an idealization of the full data input (i.e., they are a band pass version). Thus, they have lower values of the intrinsic Doppler width and higher spikes in the range-Doppler correlation. Think of this as a side effect of the fact that the real data may have a broad frequency range for the ocean wave response; but the Focus3D model only models the dominant swell.

Also, note that the upper-right section of Figure 22 shows that the intrinsic Doppler width versus time for **SANKO**. The two curves for Data and Out are nicely in sync. This is due to a method in the algorithm that filters the covariance data to model as a narrow band process but not a pure harmonic. I found that the use of a pure harmonic at the estimated ocean wave frequency does not work well because it gets out of sync with the data at one point or another in time. The solution is to use a slowly varying wave period. This is implemented via a real space filter based on chapeau functions, but any reasonable band-pass filter would probably work as well.

## 13. Summary

Focus3D is a new version of the Global Motion Model from the Melendez-Bennett paper [1] that is based on a physical model of a ship's motion that estimates variations in both aspect and tilt angles.

The program was motivated by the need for advanced physical analysis for ISAR data to support ATR algorithms. The code does this by:

1. Identifying the Profile and Plan frames and providing individual and composite versions of them by scaling the original range-Doppler image to meters of range and cross-range in the drydock coordinate system.

2. Implementing a robust ship length estimation algorithm. Length itself is a strong indicator of ship type.

3. Identifying the first range and last range cells for the ship. The time variance for the first cell is normally smaller than the variance for the last cell since the last cell has errors due to shadowing and multipath effects. Thus, it would seem to be reasonable for an ATR algorithm to measure features relative to the minimum range of the ship.

4. Automatically detecting times when the rigid body logic is flawed by multiple targets in the scene or strong interference.

This paper provides some of the data that validate the Focus3D algorithm for simulated data, for some classical data sets, and for selected high-quality cases from the MSR2. This data is from a wide variety of commercial ships, and it has strong auxiliary data, including AIS recordings. The AIS data was particularly vital both because it gave the ship ID and because it gave real-time measurements of the ship course. The accurate course measurements allowed the analysis program to verify that most of the courses were, as expected, nearly constant; but one case involved a sharp turn. The knowledge of the actual course also meant that the mean aspect angles used in the length estimates are more accurate than the tracker estimates from the system.

Taken together these examples are still only a small sample of the data analysis that has supported the development of Focus3D. This development began in 1994 with a contract from to develop an ISAR algorithm for an important radar. Further work was done sporadically over the last 28 years as employment responsibilities and contract jobs varied. However, the basic theme here is the same – Focus3D is an attempt to derive as much information as possible from the data from existing or easily built ISAR systems. Many of the algorithms in the basic ISAR processor have been incorporated into radars that are flying now. However, the exploitation of 3-D information has just begun. The purpose of these papers is to show that the 3-D algorithms are straightforward to implement and that they produce results that are useful for ATR algorithms. I encourage ATR developers to consider testing these 3-D methods to identify the Profile and Plan frame views, to improve the LOA and Rmin/Rmax estimates, and to identify the times during the ISAR dwells when the data has problems due to extra ships in the scene or interference.

## Biographical Summary

John R. Bennett received a PhD in Meteorology from the University of Wisconsin in 1972. For the next 13 years he did research at NOAA and taught oceanography at MIT. In 1986 he joined the Environmental Research Institute of Michigan to work on radar measurements of ocean waves and water depth, with some experience on ocean surface targets. For 24 years starting in 1991 he was Senior Scientist at SAIC/Leidos in San Diego, concentrating on radar and lidar signal processing. In this position he wrote and tested the initial ISAR processors for several important radars. From January 2015 to June 2022, he was Chief Scientist at RDRTec where he worked on ISAR processing. At present he is semi-retired but is working to see 3-D ISAR algorithms applied to real-world radars.

# Figure 1 (See caption list)

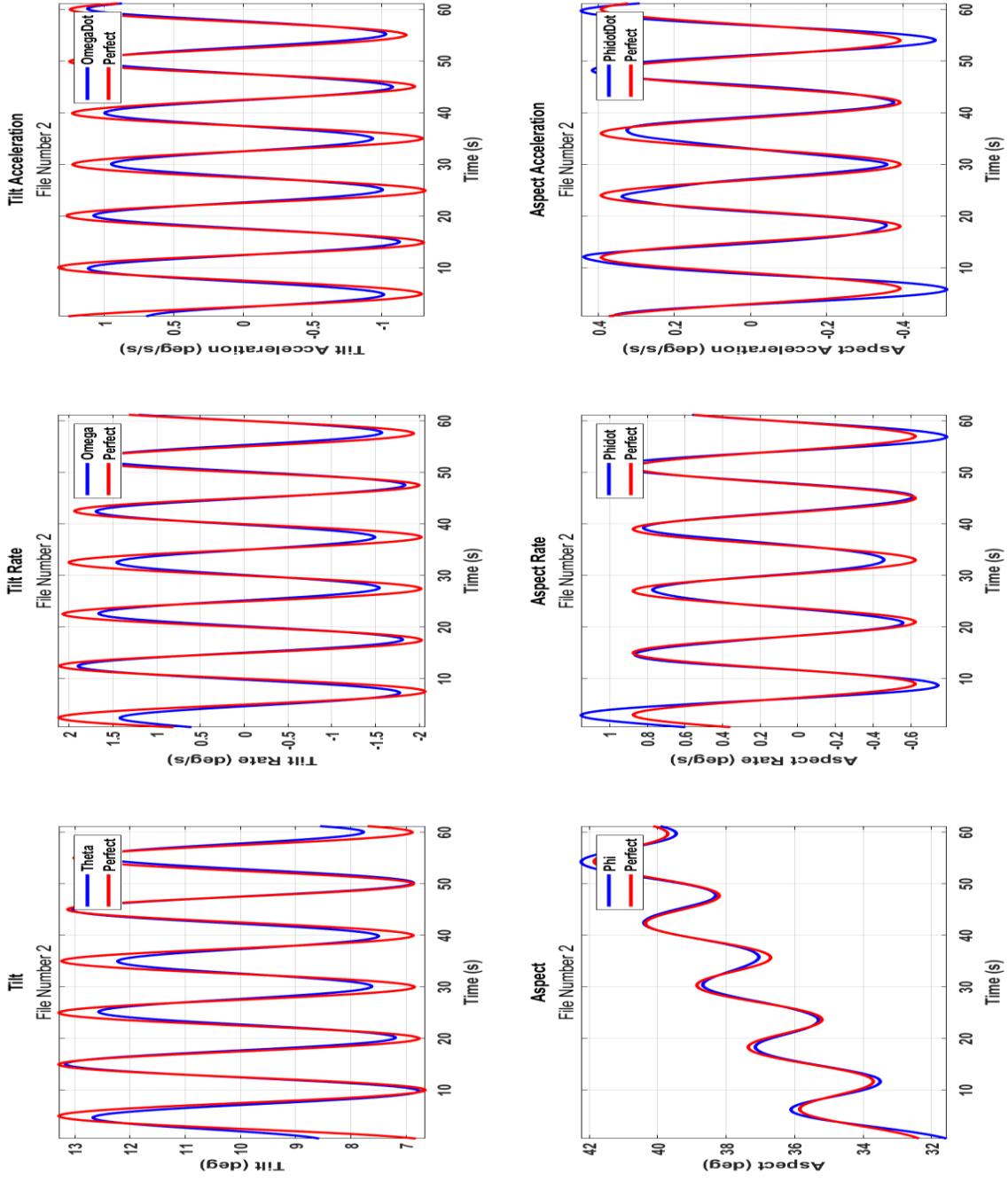

**Figure 2 (See caption list)**

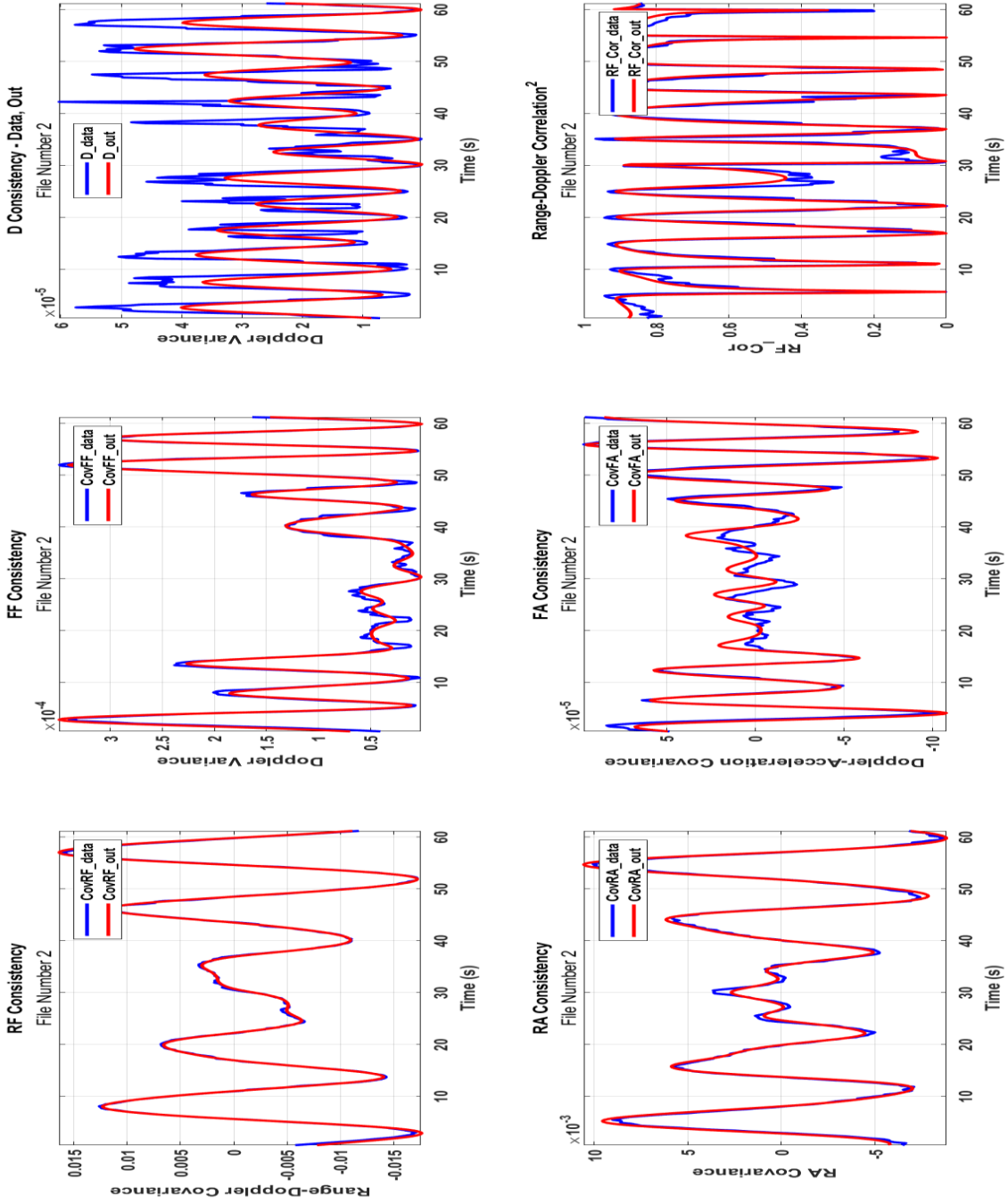

**Figure 3 (See caption list)**

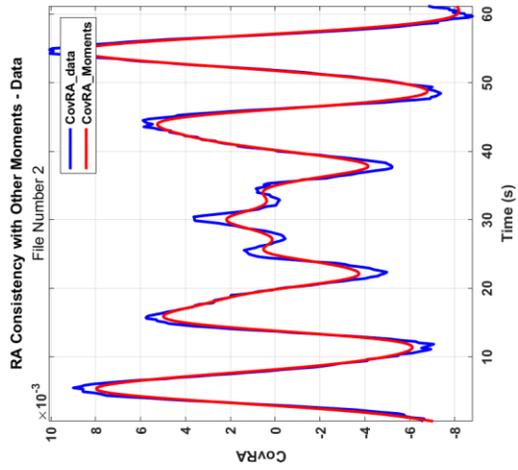
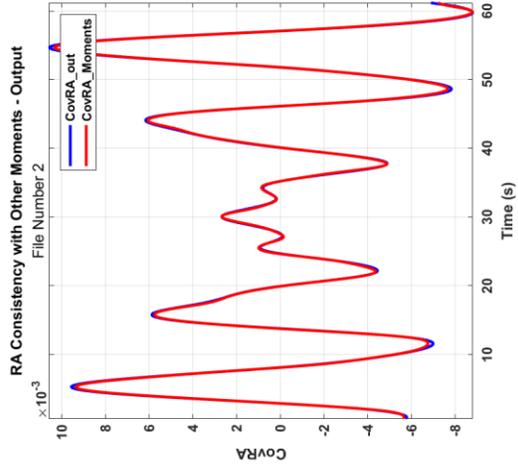
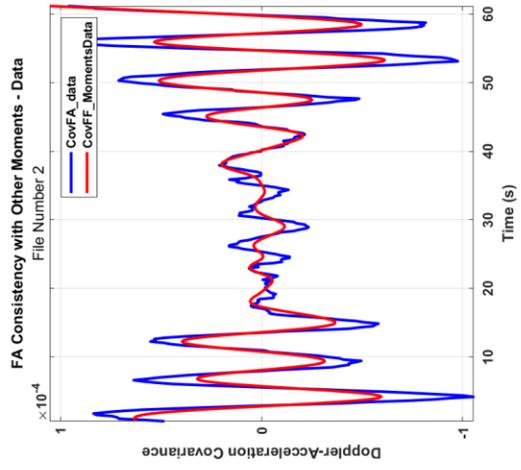
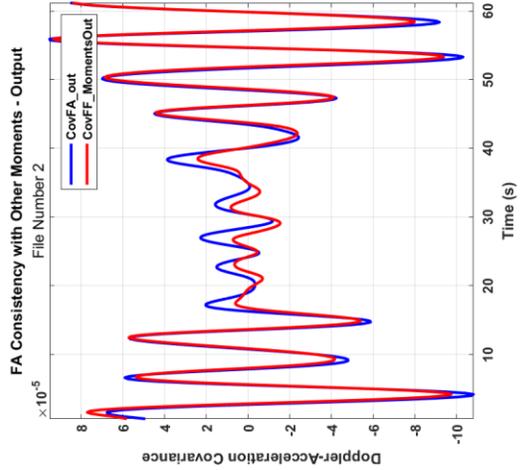

**Figure 4 (See caption list)**

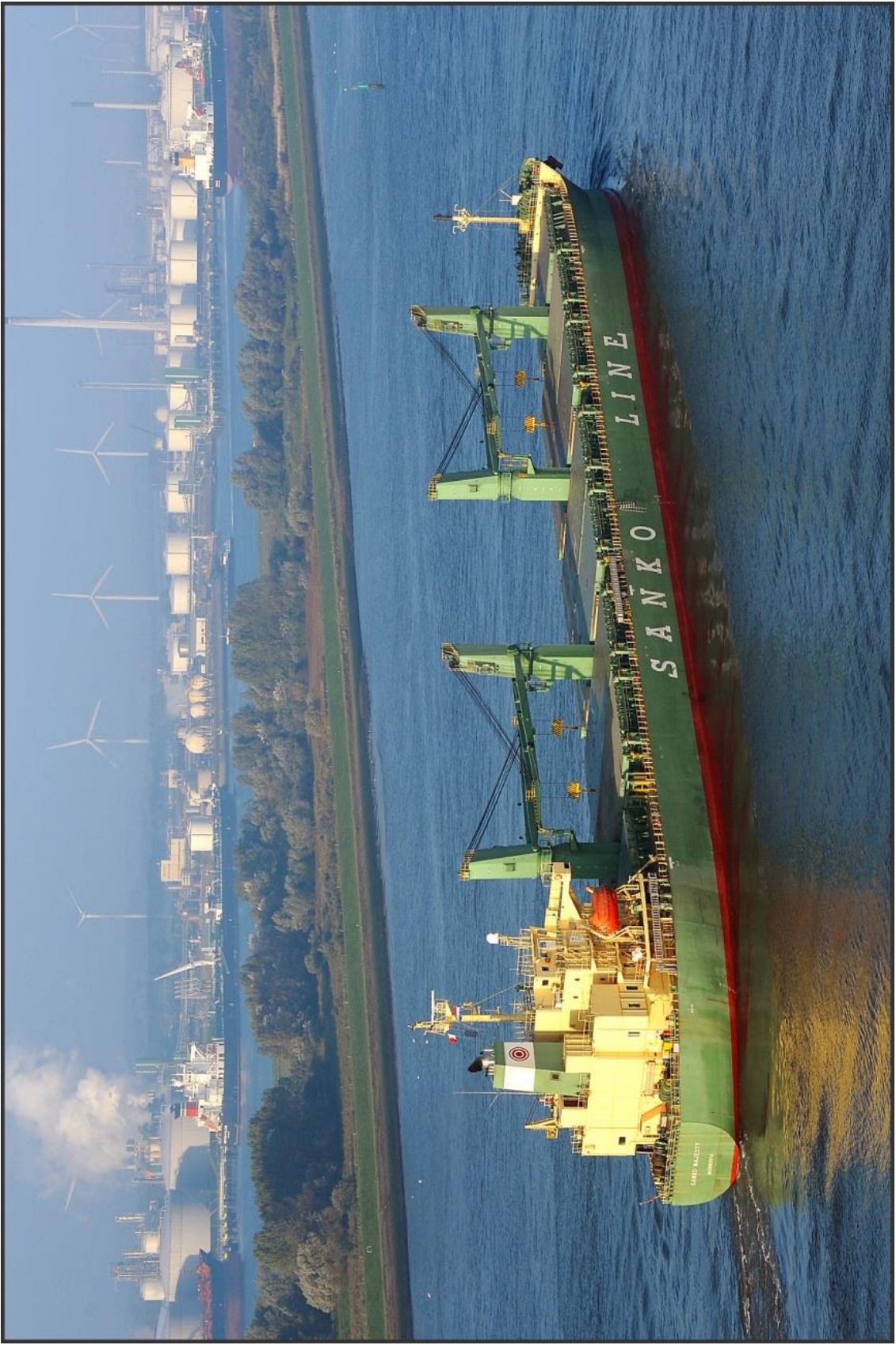

# Figure 5 (See caption list)

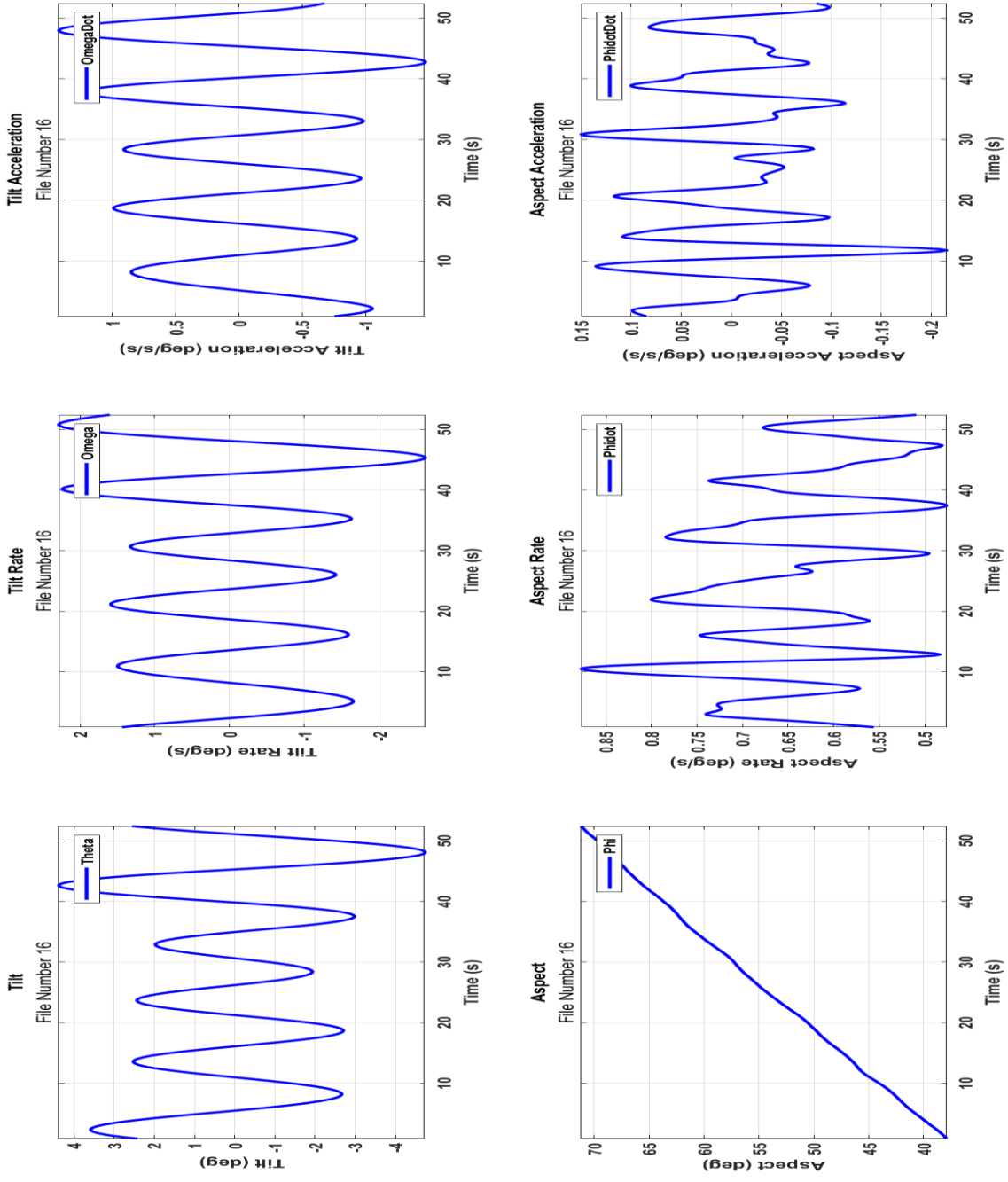

**Figure 6 (See caption list)**

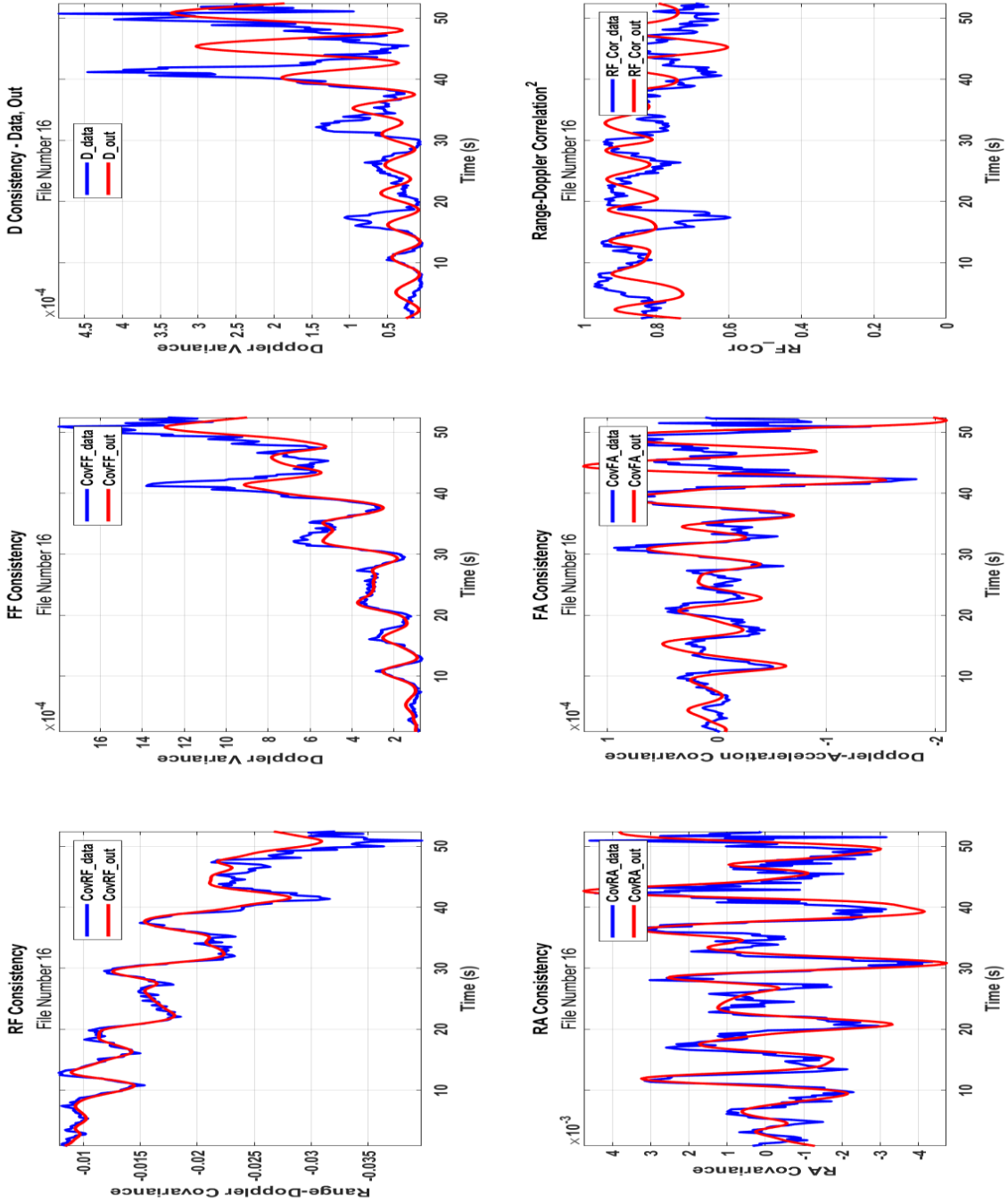

**Figure 7 (See caption list)**

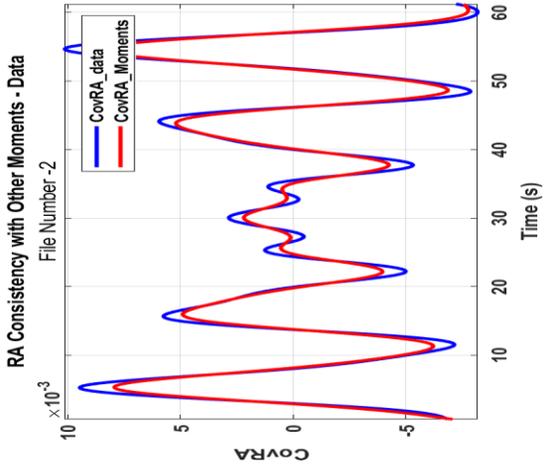
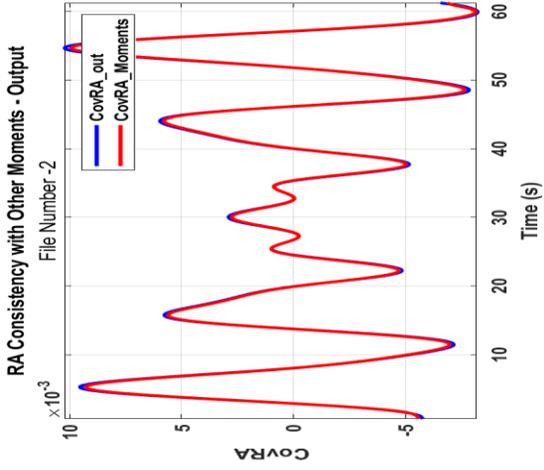
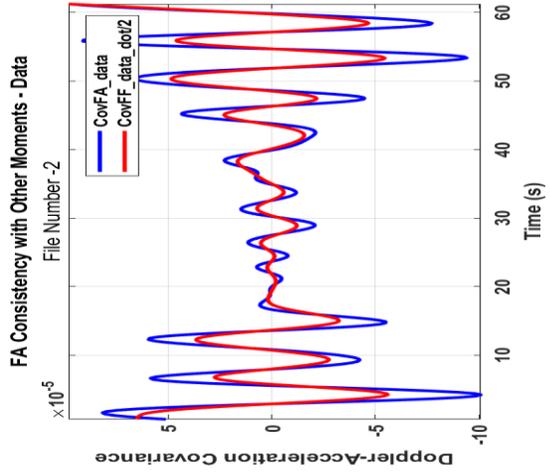
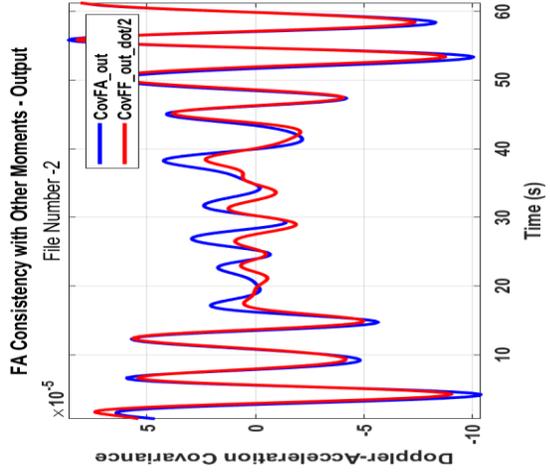

# Figure 8 (See caption list)

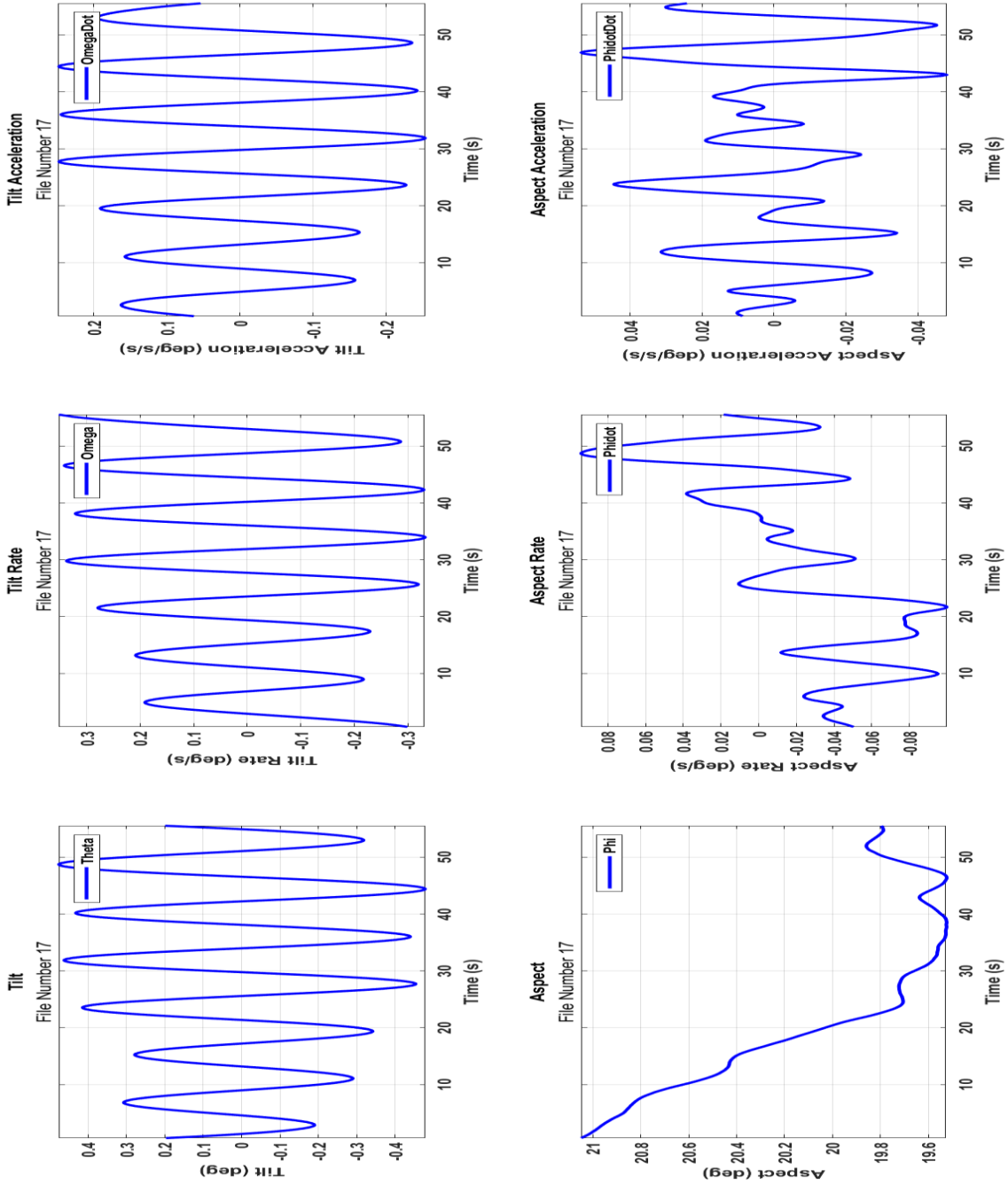

**Figure 9 (See caption list)**

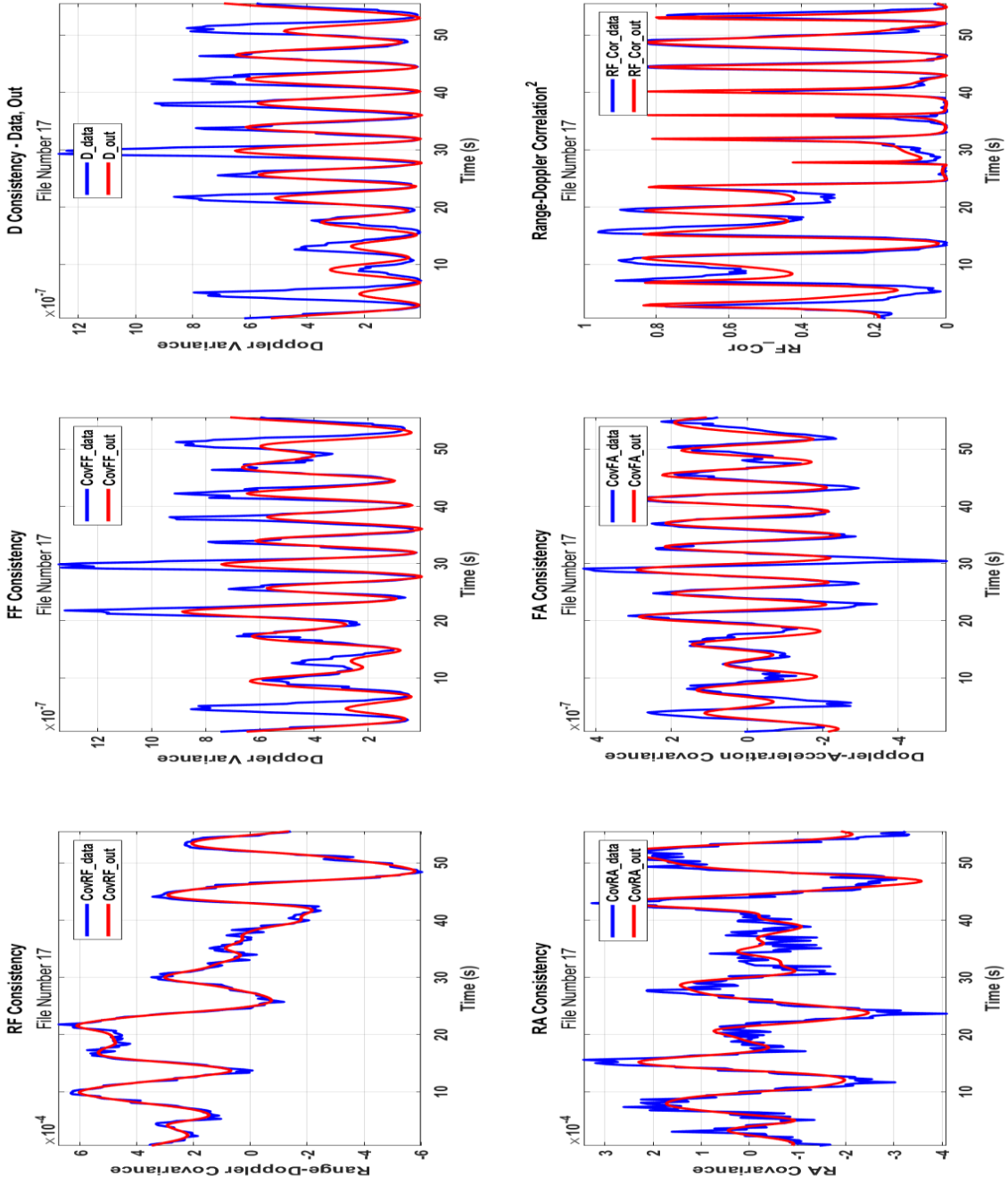

**Figure 10 (See caption list)**

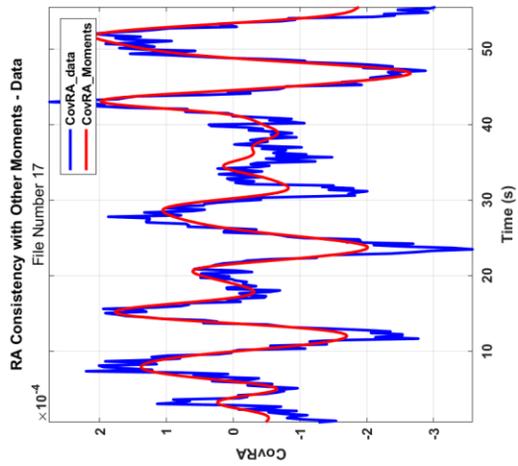
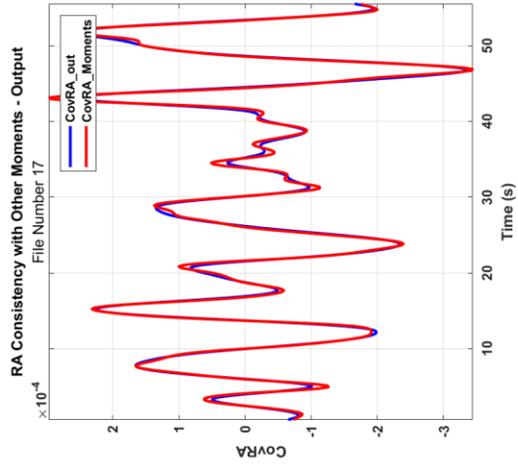
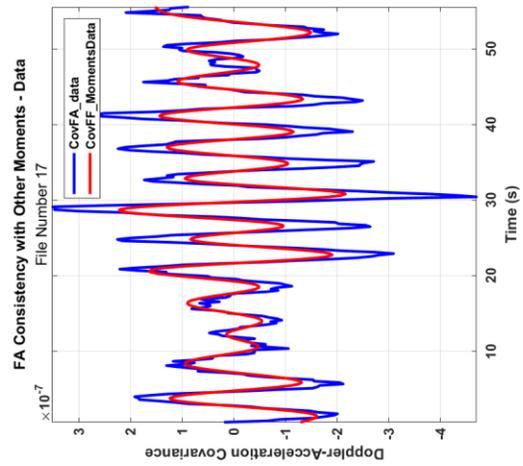
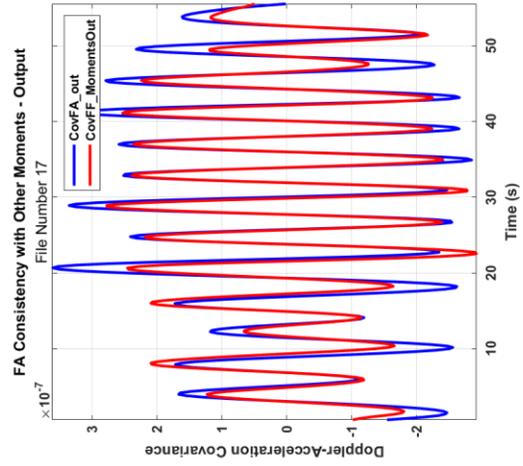

Figure 11 (See caption list)

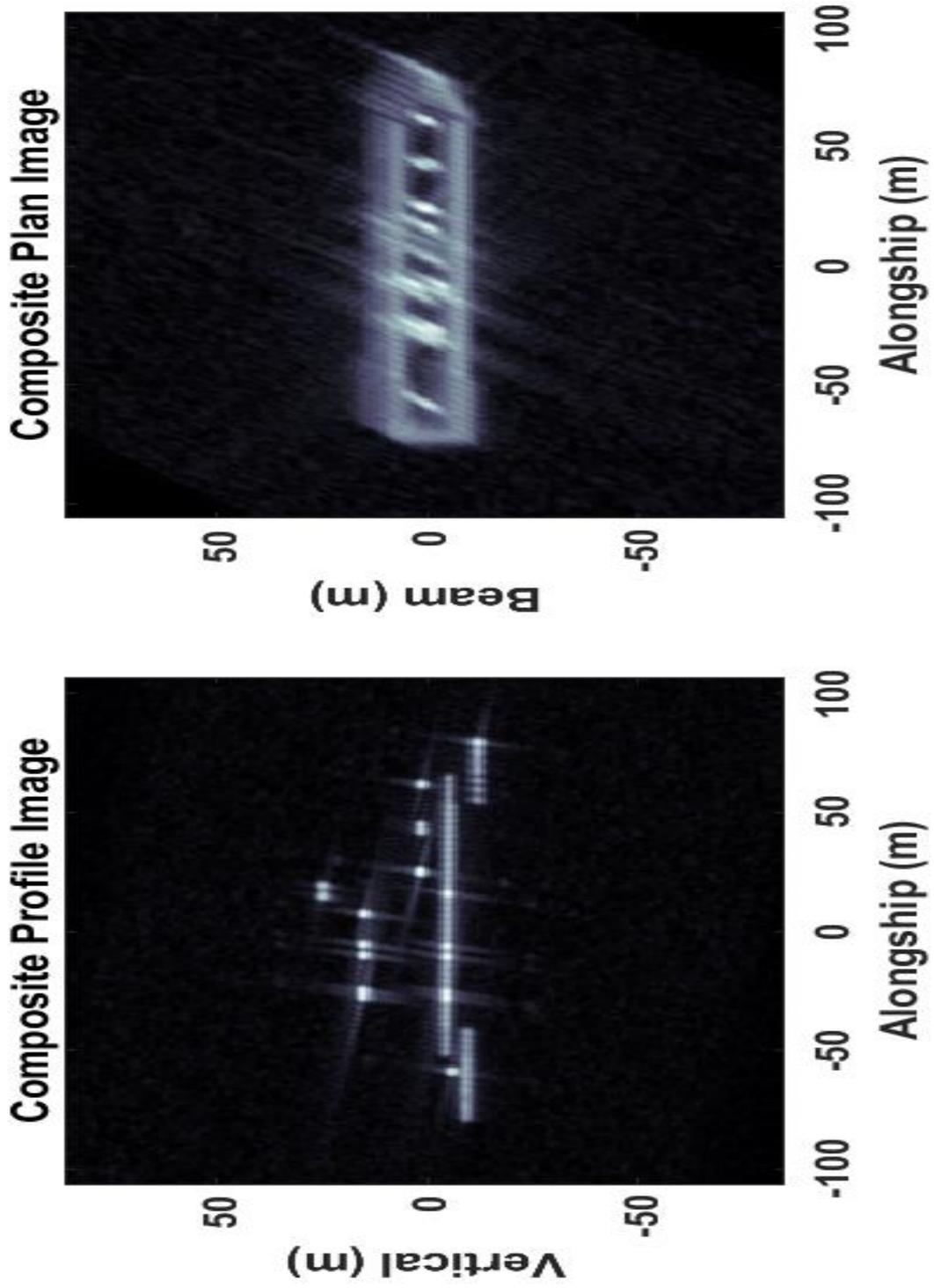

**Figure 12 (See caption list)**

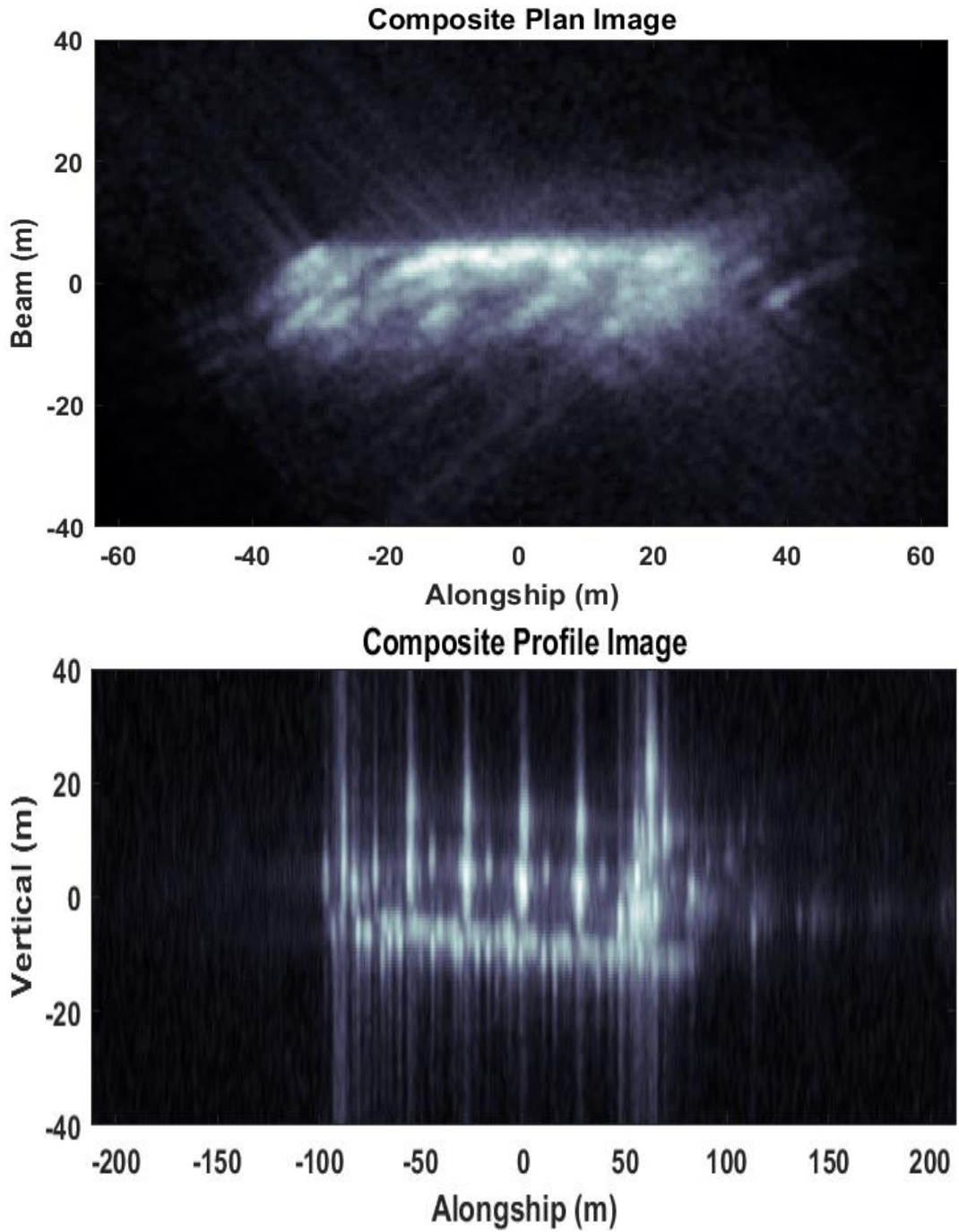

**Figure 13 (See caption list)**

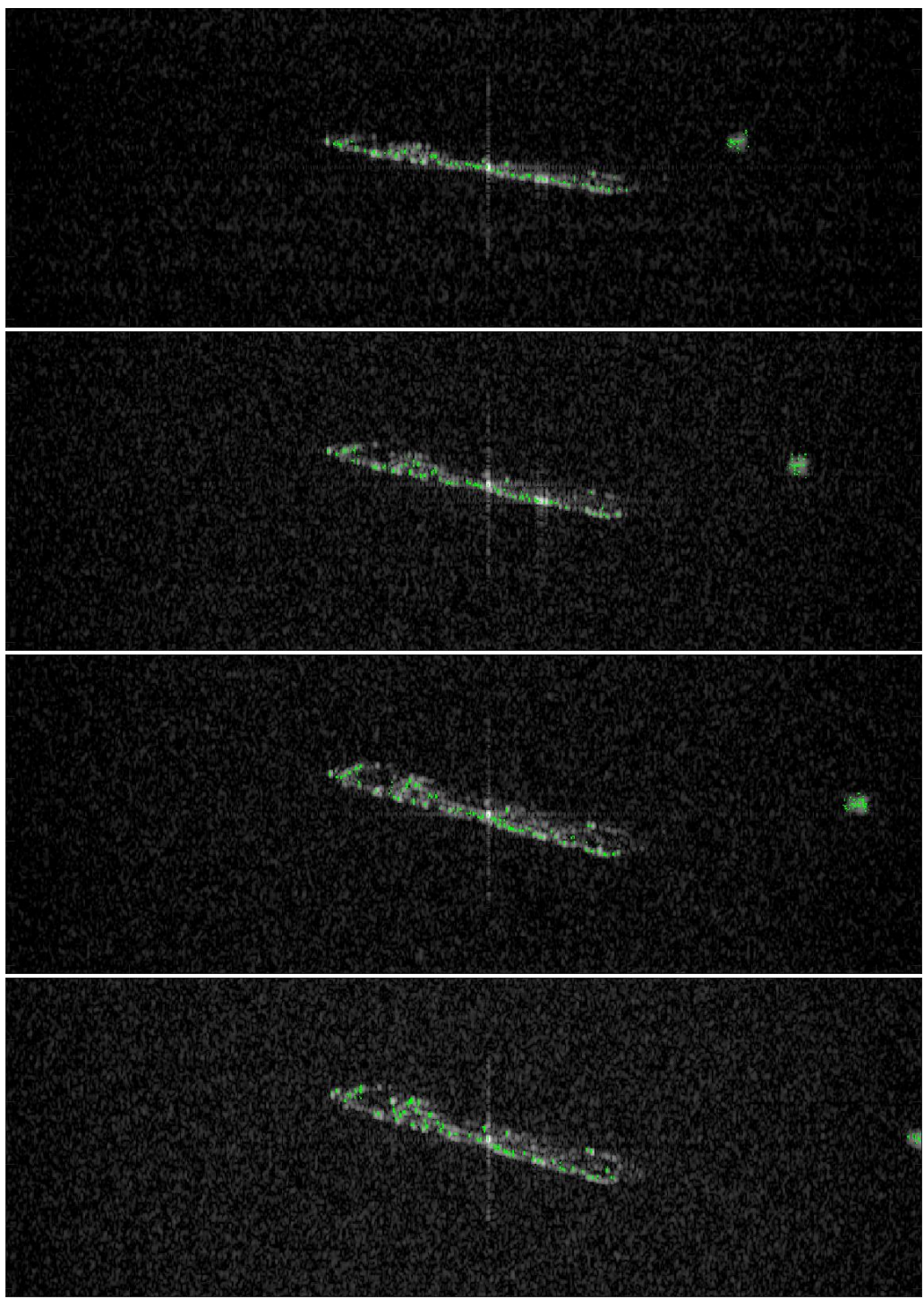

**Figure 14 (See caption list)**

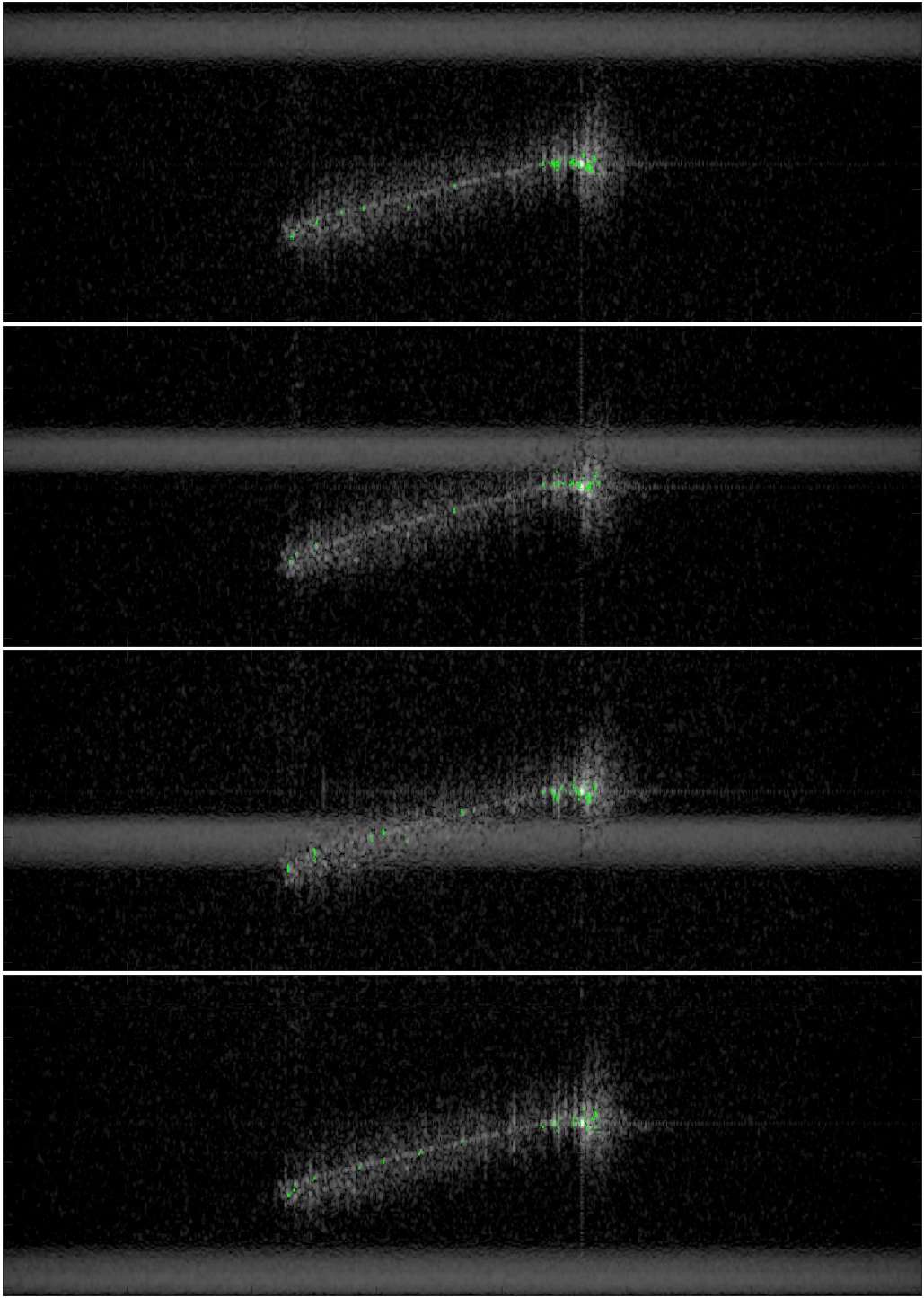

**Figure 15 (See caption list)**

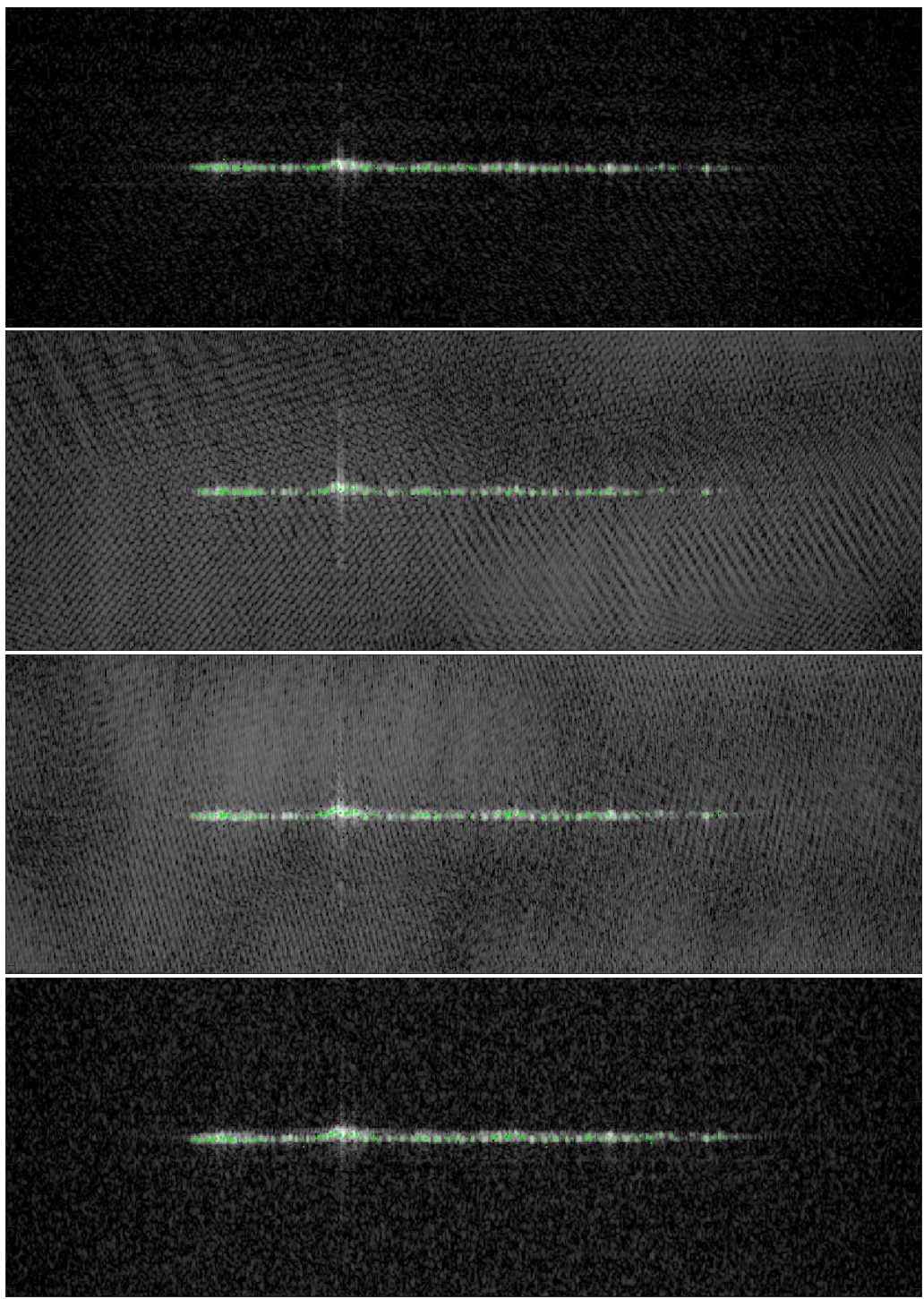

**Figure 16 – (See Captions list)**

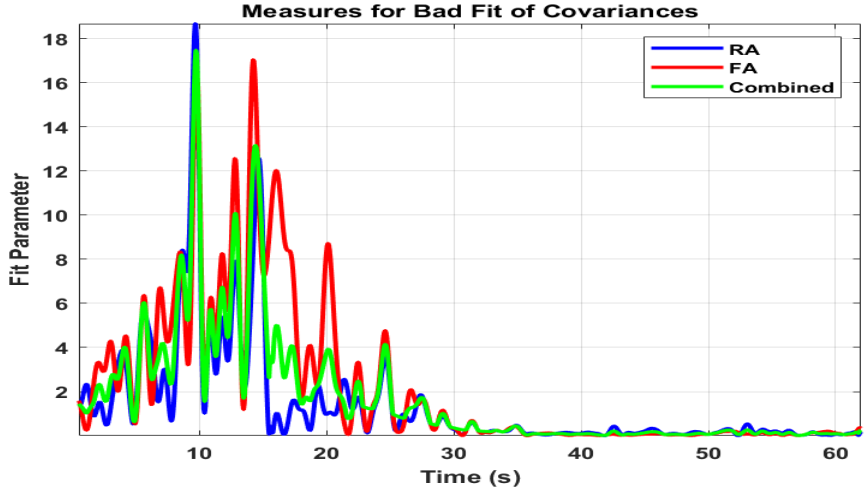

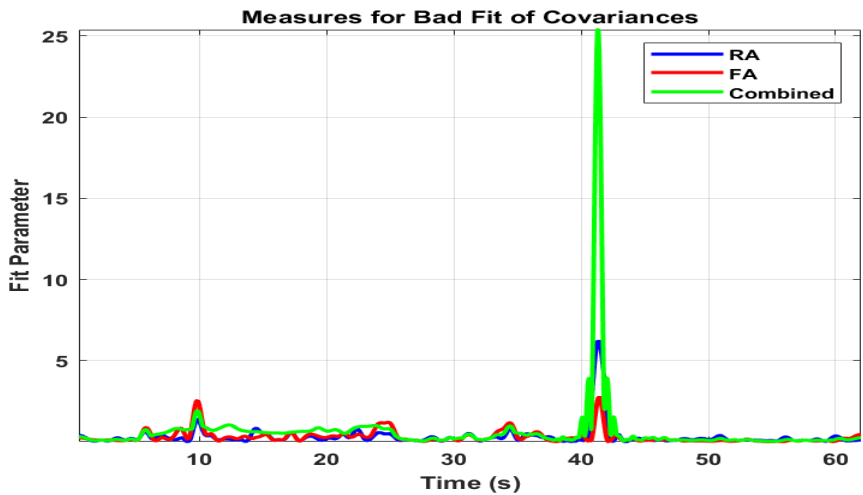

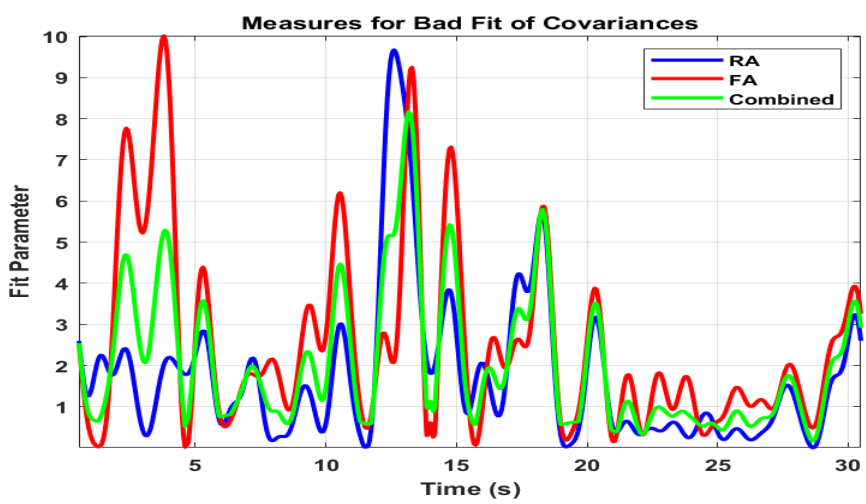

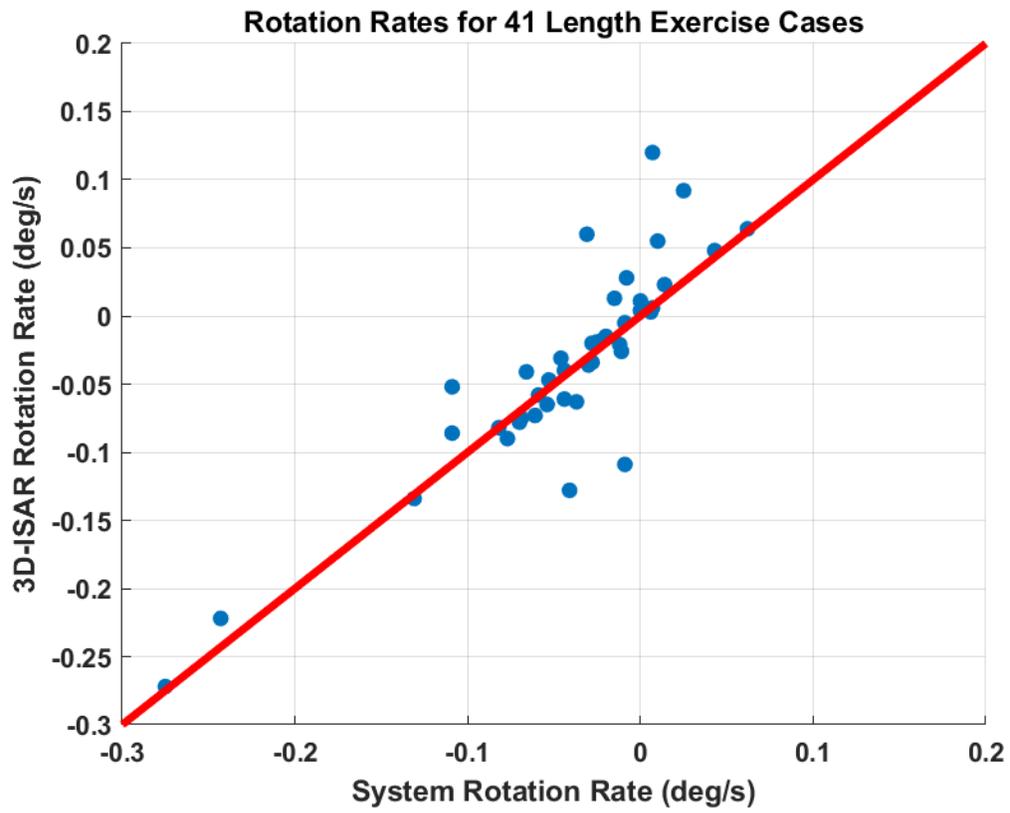

Figure 17: Apparent rotation rate - Aux vs 3-D ISAR

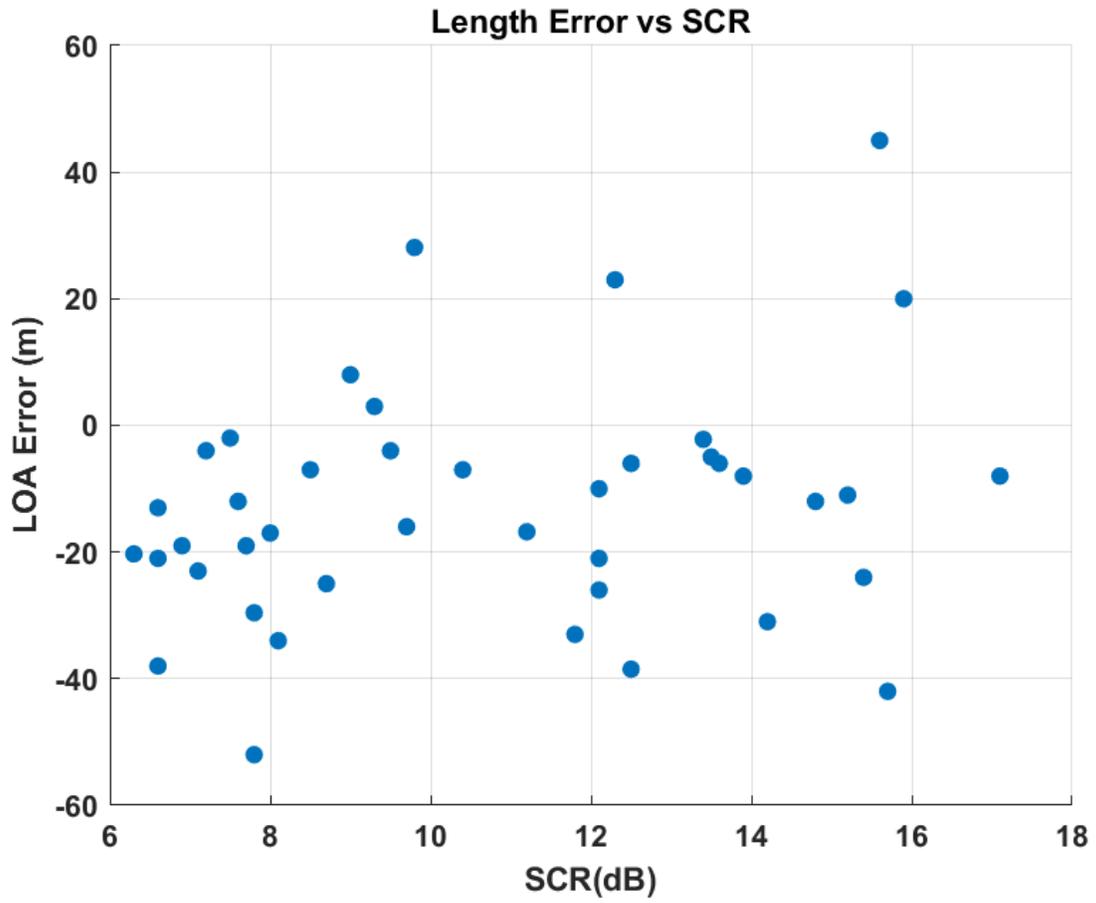

Figure 18: Ship Length error versus Signal-to-Clutter Ratio for 43 MSR2 cases

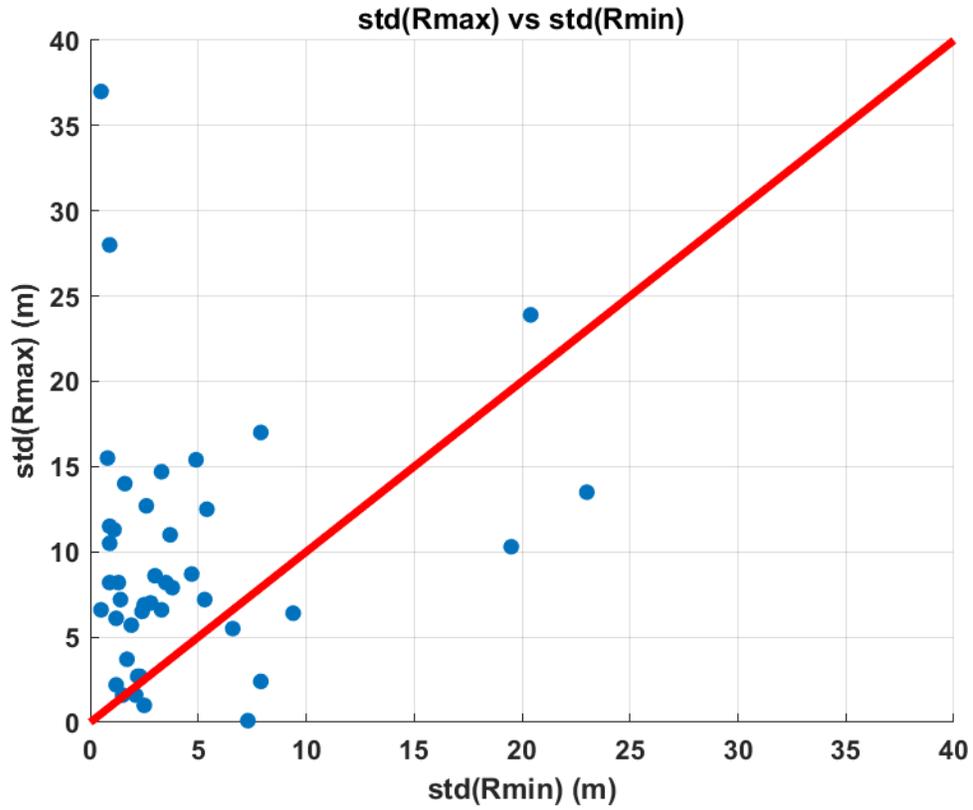

Figure 19: Standard deviations over time of the min and max ranges for 43 cases

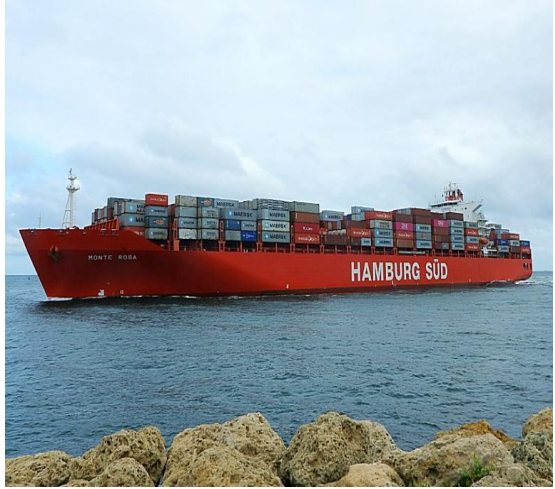

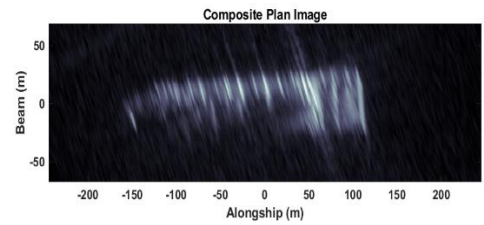

**Figure 20: MONTE ROSA (272 m container ship), with composite plan view image**

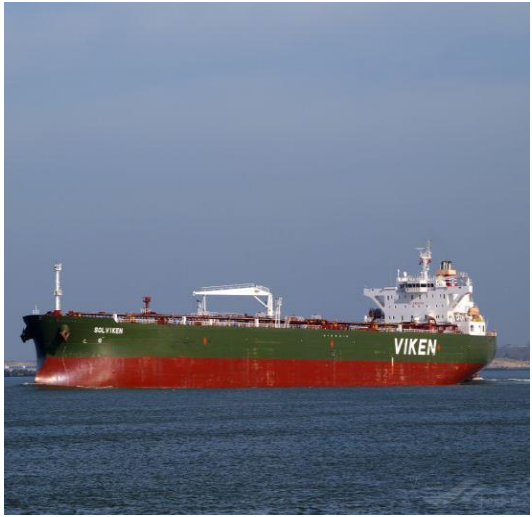
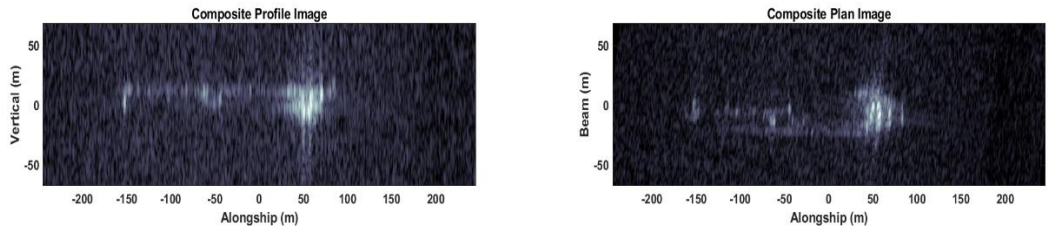

**Figure 21: SOLVIKEN (249 m tanker), with composite profile and plan view images**

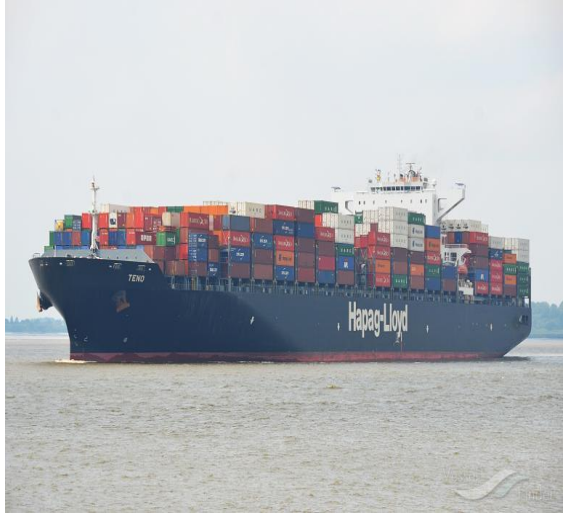
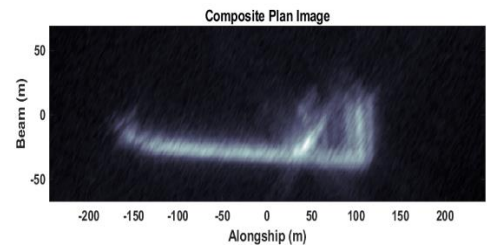

**Figure 22: TENO (299 m container ship), with composite plan view image**

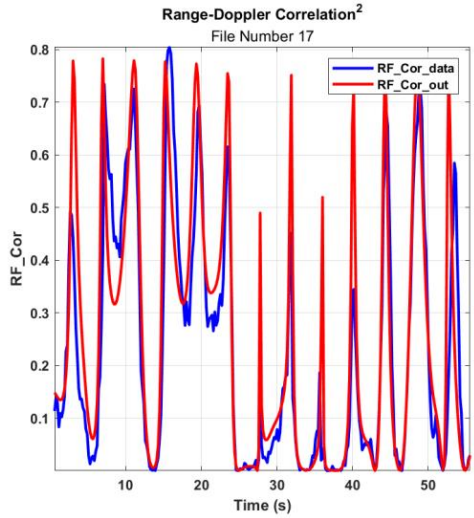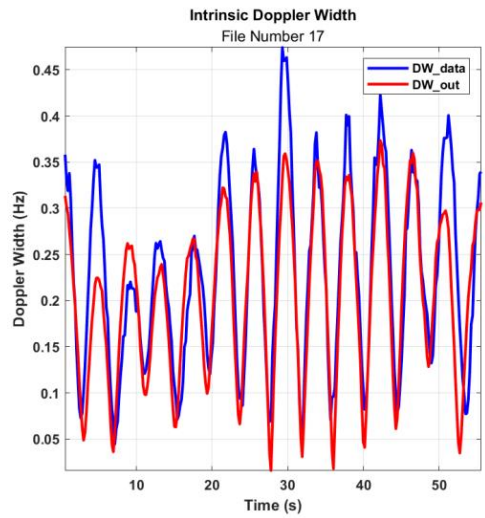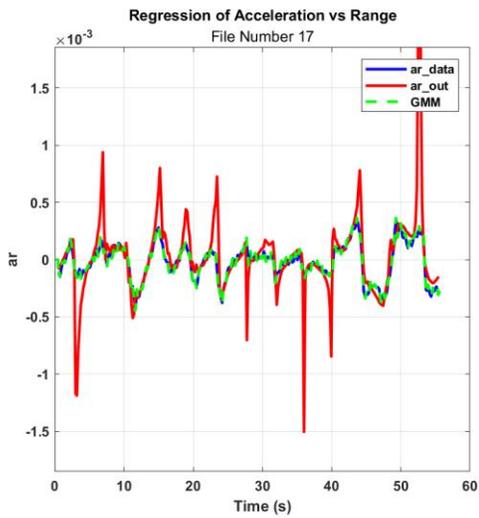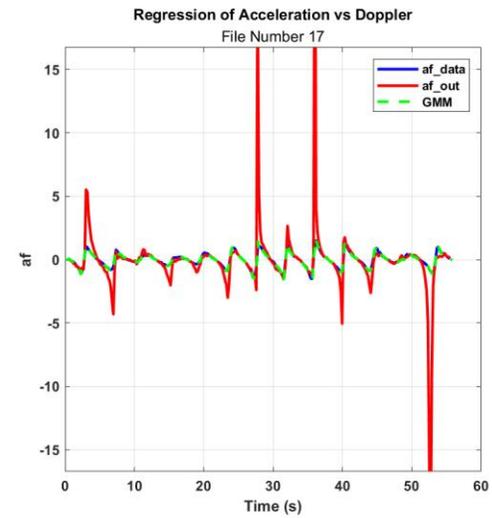

**Figure 23: Analysis of focus parameters and range-Doppler correlation for SANKO**

# Figure Captions

1. Angle results for a simulated data case compared to the exact solution

2. Moments of the elementary statistics – comparison of the moments of the input data and the moments derived from the output results

3. Consistency checks using the acceleration moments (the covariances of Range-Acceleration and Doppler-Acceleration), data versus output moments

4. The SANKO MAJESTY, a bulk carrier with cranes, like the SANKO ship of opportunity discussed here

5. Angle results for 'Large Ship Turning'

6. Large Ship Turning - Moments of the elementary statistics – comparison of the moments of the input data and the moments derived from the output results

7. Large Ship Turning - Consistency checks using the acceleration moments (the covariances of Range-Acceleration and Doppler-Acceleration), data versus output moments

8. Angle results for SANKO, a bulk carrier with large cranes

9. SANKO - Moments of the elementary statistics – comparison of the moments of the input data and the moments derived from the output results

10. SANKO - Consistency checks using the acceleration moments (the covariances of Range-Acceleration and Doppler-Acceleration), data versus output moments

11. Composite profile and plan images for simulated data

12. Top – composite plan view of Large Ship Turning. Bottom – composite profile view of SANKO.

13. Sequence of ISAR movie frames (top to bottom) showing an extra ship target that migrates out the far range over time

# Appendix: Equations for the 3-D ISAR Algorithm

In the original implementation Focus3D Algorithm assumed that the grazing angle was small, and that the azimuth motion was slower than the tilt motion. These assumptions have been relaxed and the present algorithm uses a more general formulation. This document summarizes the current general version of the algorithm.

Assumptions for the current version of Focus3D:

1. The fundamental assumption is that the radar can only sense two degrees of rotation – roll about the radar line-of-sight cannot be detected. A solid body has a third degree of rotation but unless the angular change over time is large a radar cannot estimate or exploit the third dimension. The two angles that the radar can measure are Azimuth ($\phi$) and Tilt ($\theta$).
2. The grazing angle can be any value but in practice any angle close to straight down would not give good imaging performance – we expect to operate in the range of zero (low altitude) to 45 degrees.
3. The range is assumed large compared to the ship's dimensions – this is a good assumption, and it allows us to ignore wavefront curvature.
4. We assume that the ship is under track and that thus we have an estimate of its location and heading – from this and knowledge of the radar's position we can get a good estimate of the mean azimuth and tilt angles during the imaging.
5. The ship is assumed to be 'thin' or at least not fat – this allows the ratio of the height or beam to the length to be expanded in terms of the mean and one additional term. This assumption is less accurate for small fat boats like a tugboat. It is a good assumption for a large commercial ship and for most military ships. In the 'thin ship' approximation we can also estimate the alongship coordinate and length from the range coordinate and the cosines of the mean grazing and azimuth angles.
6. In the derivation we approximate the tangent function with only the mean value and the first derivative.

There are two major elements to the 3-D mathematics discussion:

A.1 Estimation of the time-dependent values of the azimuth and tilt angles

A.2 Exploitation of these results to estimate the shape of the ship in its natural (or 'drydock') coordinates (x = alongship, y = beam, z = height)

## A.1. Estimation of the Azimuth and Tilt angles

The 3D code develops its estimates of azimuth and tilt angles from measurements of range, range rate and range acceleration produced by the ISAR processor. The ISAR processor identifies as many point-like scatterers as it can to focus the ISAR imagery. In this process it produces a statistical/physical model that we call the Global Motion Model. But the 3-D code does its own analysis of these raw target reports.

The basic model calculates the aspect and tilt angles from elementary statistics of the target reports on the range, range rate, and range acceleration. The most important statistic is the range-Doppler covariance function. Other significant statistics used are the Doppler variance and the Doppler skewness (third moment).

All the equations below apply to each scatterer identified by the ISAR processor – in other words, each equation here is a family of equations. The expected values here are thus based on the sum over the scatterers identified for each time frame, a number that varies with time as the SNR values for each scatterer change or they become shadowed by other features.

When calculating the time derivatives, the only variables that are functions of time are the two angles. The XYZ (subscript zero) coordinates are the unknown values of the coordinates in the natural (drydock) coordinate system and are assumed to be independent of time.

Define the two degrees of rotation with respect to the radar as:

$$Aspect\ Angle = \phi(t)$$

$$Tilt\ Angle = \theta(t)$$

*Equation 1 – Definition of range*
$$r = x_0 * \cos\theta * \cos\varphi - y_0 * \cos\theta * \sin\varphi - z_0 * \sin\theta$$

*Equation 2 – Derivative to get range-rate*
$$\dot{r} = x_0 * \left(-\dot{\theta} * \sin\theta * \cos\varphi - \dot{\varphi} * \cos\theta * \sin\varphi\right)$$
$$+ y_0 * \left(\dot{\theta} * \sin\theta * \sin\varphi - \dot{\varphi} * \cos\theta * \cos\varphi\right)$$
$$- z_0 * \dot{\theta} * \cos\theta$$

We need to compute
$$\frac{\langle r * \dot{r}\rangle}{\langle r^2\rangle}, \quad \frac{\langle \dot{r}^2\rangle}{\langle r^2\rangle}$$

We ignore cross product terms based on simulations and logic that shows that they should not be significant:

$$\langle x_0 * y_0 \rangle = \langle x_0 * z_0 \rangle = \langle y_0 * z_0 \rangle = 0$$

*Equation 3 – Range-Doppler Covariance*

$$\langle r * \dot{r} \rangle = \langle x_0^2 \rangle * \cos\theta \cos\varphi * (-\dot{\theta} * \sin\theta * \cos\varphi - \dot{\varphi} * \cos\theta * \sin\varphi)$$
$$+ \langle y_0^2 \rangle * \cos\theta \sin\varphi * (-\dot{\theta} * \sin\theta * \sin\varphi + \dot{\varphi} * \cos\theta * \cos\varphi)$$
$$+ \langle z_0^2 \rangle * \dot{\theta} * \cos\theta * \sin\theta$$

*Equation 4 – Range variance*

$$\langle r^2 \rangle = \langle x_0^2 \rangle * \cos^2\theta * \cos^2\varphi + \langle y_0^2 \rangle * \cos^2\theta * \sin^2\varphi + \langle z_0^2 \rangle * \sin^2\theta$$

*Equation 5 – Doppler (range-rate) variance*

$$\langle \dot{r}^2 \rangle = \langle x_0^2 \rangle * (\dot{\theta} * \sin\theta * \cos\varphi + \dot{\varphi} * \cos\theta * \sin\varphi)^2$$
$$+ \langle y_0^2 \rangle * (\dot{\theta} * \sin\theta * \sin\varphi - \dot{\varphi} * \cos\theta * \cos\varphi)^2$$
$$+ \langle z_0^2 \rangle * (\dot{\theta} * \cos\theta)^2$$

$$Divide \quad \langle r * \dot{r} \rangle, \langle r^2 \rangle, \langle \dot{r}^2 \rangle \quad by \quad \langle x_0^2 \rangle * \cos^2\theta * \cos^2\varphi$$

*Equation 6*

$$Define \quad bsq = \frac{\langle y_0^2 \rangle}{\langle x_0^2 \rangle}, \quad hsq = \frac{\langle z_0^2 \rangle}{\langle x_0^2 \rangle}$$

Equation 7

$$\frac{\langle r^2 \rangle}{\langle x_0^2 \rangle * \cos^2\theta * \cos^2\varphi} = 1 + bsq * \tan^2\varphi + hsq * \frac{\tan^2\theta}{\cos^2\varphi}$$

*Equation 8*

$$\frac{\langle r^2 \rangle}{\langle x_0^2 \rangle * \cos^2\theta * \cos^2\varphi}$$
$$= (\dot{\varphi} * \tan\varphi + \dot{\theta} * \tan\theta)^2 + bsq * (\dot{\varphi} - \dot{\theta} * \tan\theta * \tan\varphi)^2 + hsq * \frac{\dot{\theta}^2}{\cos^2\varphi}$$

*Equation 9 – Approximate range-Doppler Regression Slope*

$$\frac{\langle r * \dot{r} \rangle}{\langle x_0^2 \rangle * \cos^2\theta * \cos^2\varphi} = -\dot{\varphi} * \tan\varphi - \dot{\theta} * \tan\theta$$
$$+ bsq * (\dot{\varphi} * \tan\varphi - \dot{\theta} * \tan\theta * \tan^2\varphi)$$
$$+ hsq * \dot{\theta} * \frac{\tan\theta}{\cos^2\varphi}$$

*Equation 10*

$$CovRF = \frac{\langle r * \dot{r} \rangle}{\langle r^2 \rangle} = \left. \frac{\langle r * \dot{r} \rangle}{\langle x_0^2 \rangle * \cos^2\theta * \cos^2\varphi} \middle/ \frac{\langle r^2 \rangle}{\langle x_0^2 \rangle * \cos^2\theta * \cos^2\varphi} \right.$$

*Equation 11*

$$Define \quad denom = \frac{\langle r^2 \rangle}{\langle x_0^2 \rangle * \cos^2\theta * \cos^2\varphi} = 1 + bsq * \tan^2\varphi + hsq * \frac{\tan^2\theta}{\cos^2\varphi}$$

Substituting Equation 9 and Equation 11 into Equation 10 yields:

*Equation 12*

$$-CovRF * denom = \dot{\varphi} * \tan\varphi + \dot{\theta} * \tan\theta - bsq * (\dot{\varphi} * \tan\varphi - \dot{\theta} * \tan\theta * \tan^2\varphi)$$
$$-hsq * \dot{\theta} * \frac{\tan\theta}{\cos^2\varphi}$$

Which reduces to:

*Equation 13*

$$-CovRF * denom = \dot{\varphi} * (1 - bsq) * \tan\varphi + \dot{\theta} * \left(1 + bsq * \tan^2\varphi - \frac{hsq}{\cos^2\varphi}\right) * \tan\theta$$

Now we use the "thin ship" approximation where we assume that angle fluctuations can be approximated by the mean angles when multiplied by bsq and hsq.

NOTE: For simplicity of notation redefining denom with subscript zero on angles

$$Define, \quad denom = 1 + bsq * \tan^2\varphi_0 + hsq * \frac{\tan^2\theta_0}{\cos^2\varphi_0}$$
$$define \; P = 1 - bsq$$
$$define \; Q = 1 + bsq * \tan^2\varphi_0 - \frac{hsq}{\cos^2\varphi_0}$$

*Equation 14*

$$-CovRF * denom = P * \tan\varphi * \dot{\varphi} + Q * \tan\theta * \dot{\theta}$$

$$Let, \quad LHS = Data \; Measurement$$

$$LHS = \overline{LHS} + LHS' = -CovRF * denom$$

Remove "Mean $\varphi$" term, $\varphi_M$ Candidates:

$$\varphi_L = \varphi_0 + \frac{\overline{LHS} * (t - \bar{t})}{P * \tan(\varphi_0)}, \quad Linear$$
$$\varphi_2 = \varphi_2 \dots \quad Two \; Terms$$

$$\varphi_A = \varphi_0 + Zero\ Mean\left(\dot\varphi = \frac{\overline{LHS}}{P*\tan(\varphi_0)}\right), \quad All\ (the\ choice\ implemented)$$

$$LHS = P*(\tan\varphi_0 + \varphi'*\sec^2\varphi_0)*\dot\varphi' + Q*(\tan\theta_0 + \theta'*\sec^2\theta_0)*\dot\theta'$$
$$where,\ \varphi' = \varphi - \varphi_0,\ \theta' = \theta - \theta_0$$

Integrate in time:

$$\int LHS\ dt = \overline{LHS}(t-\bar t) + \int LHS'dt + T, \quad where\ T = Intergtation\ constant$$

$$\int LHS\ dt = P*\left(\varphi'*\tan\varphi_0 + \frac{\varphi'^2}{2}*\sec^2\varphi_0\right) + Q*\left(\theta'*\tan\theta_0 + \frac{\theta'^2}{2}*\sec^2\theta_0\right)$$

Define $\varphi_M$ as the solution to:

$$\overline{LHS}(t-\bar t) = P*\varphi_M*\tan\varphi_0 + \frac{P}{2}\sec^2\varphi_0*\varphi_M{}^2$$

Subtract out $\varphi_M$ => $\varphi' = \hat\varphi + \varphi_M$

$$P\left[\varphi'\tan\varphi_0 + \sec^2\varphi_0\frac{\varphi'^2}{2}\right]$$

$$= P\left[(\hat\varphi + \varphi_M)\tan\varphi_0 + \sec^2\varphi_0\frac{(\hat\varphi+\varphi_M)^2}{2}\right]$$

$$= P[\hat\varphi\tan\varphi_0 + \sec^2\varphi_0*\varphi_M\hat\varphi + \sec^2\varphi_0\frac{\hat\varphi^2}{2}]$$

But, $= \tan\varphi_0 + \sec^2\varphi_0*\varphi_M \doteq \tan(\varphi_M + \varphi_0)$

$$\int LHS'dt + T = P\left[\hat\varphi\tan(\varphi_M+\varphi_0) + \sec^2\varphi_0\frac{\hat\varphi^2}{2}\right] + Q[\hat\theta\tan\theta_0 + \sec^2\theta_0\frac{\hat\theta^2}{2}$$

$$A = Primary = P\hat\varphi\tan(\varphi_M+\varphi_0) + Q\hat\theta\tan\theta_0$$
$$B = Second = P\sec^2\varphi_0\frac{\hat\varphi^2}{2} + Q\sec^2\theta_0\frac{\hat\theta^2}{2}$$

$$Define, g = \frac{\hat\varphi}{\cos\varphi_0}, h = \frac{\hat\theta}{\cos\theta_0}$$
$$2B = Pg^2 + Qh^2$$
$$A = P\tan(\varphi_M+\varphi_0)\cos\varphi_0*g + Q\sin\theta_0*h$$
$$\hat P = P\tan(\varphi_M+\varphi_0)\cos\varphi_0$$
$$g = \frac{1}{\hat P}[A - Q\sin\theta_0*h]$$
$$2B = Qh^2 + \frac{P}{\hat P^2}[A^2 - 2AQ\sin\theta_0 h + Q^2\sin^2\theta_0 h^2]$$
$$h^2 = [Q + \frac{Q^2P}{\hat P^2}\sin^2\theta_0] - 2\frac{AQP}{\hat P^2}\sin\theta_0 h + \left[\frac{A^2P}{\hat P^2} - 2B\right] = 0$$

This is a quadratic equation in h with the standard mathematical notation:

$$a = Q + \frac{Q^2 P}{\hat{P}^2} \sin^2 \theta_o$$

$$b = -2 \frac{AQP}{\hat{P}^2} \sin \theta_o$$

$$c = \frac{A^2 P}{\hat{P}^2} - 2B$$

Assume $b^2 - 4ac > 0$ so that the roots are real:

$$b^2 - 4ac = \frac{4A^2 Q^2 P^2 \sin^2 \theta_o}{\hat{P}^4} - 4\left[Q + \frac{Q^2 P \sin^2 \theta_o}{\hat{P}^2}\right]\left[\frac{A^2 P}{\hat{P}^2} - 2B\right]$$

$$= \frac{4A^2 Q^2 P^2 \sin^2 \theta_o}{\hat{P}^4} + 8B\left[Q + \frac{Q^2 P \sin^2 \theta_o}{\hat{P}^2}\right] - \frac{4A^2 Q^2 P^2 \sin^2 \theta_o}{\hat{P}^4} - \frac{4QPA^2}{\hat{P}^2} > 0$$

$$\text{But,} \frac{\hat{P}^2}{P} = \frac{P^2}{P}\tan^2 \varphi_L * \cos^2 \varphi_o$$

$$\text{So, } 2B > \frac{A^2}{P * \tan^2 \varphi_L \cos^2 \varphi_o + Q * \sin^2 \theta_o}$$

This constraint is applied as a floor to B to ensure that the roots of the quadratic equation are real.

The algorithm for fitting the angles to the covariances is relatively complex but it consists of simple calculations of short time records and thus requires very small computing resources. First the algorithm estimates the dominant wave frequency by computing FFTs of the main covariances. Then it exercises the fit algorithm over a range of wave periods, normally +/-20% of the first estimate. The final wave period estimate is the one that minimizes the RMS difference between the covariance from data and the expression for the output angles.

For each test wave period the range-Doppler covariance is split into a low pass region, a wave period region, and a high pass region. For the low pass region, the program assumes that the angle fluctuations are dominated by a steady turning motion in aspect only, by integrating Equation 14 for aspect angle and applying the known mean aspect. The logic for this is that motion is strongly limited in tilt angle for time scales longer than the wave period – a ship is free to turn 360 degrees in aspect but very little in the vertical. For time scales other than the low pass region fluctuations in both angles are considered.

The process of estimating the angles for each test value of the dominant wave period is an iterative algorithm for jointly fitting the two covariances CovRF and CovFF. The code is written in terms of CovRF and the variable D, defined as CovFF-CovRF^2 because the process is simpler for these variables. During this iterative process the program makes estimates of the variables bsq and hsq defined above.

Focus3D also uses derivations like the one above for the covariances CovRA and CovFA, but these derivations are too lengthy to present here.

The result of all the mathematical approximations and assumptions is a set of equations relating the 'data' values of the 4 covariances with the 'out' versions that involve only functions of the two angles and their first two derivatives. In addition, the derivations produce 'consistency' relations that relate the acceleration covariances to the two lower order covariances and their derivatives. The program carefully monitors the differences between the data and out versions, and the consistency relations. The validity of the approximations used here has been extensively tested with simulated and real data.

## A.2. Exploitation of Angle Results to Estimate the Ship in Natural Coordinates

Define x₀(N), y₀(N), z₀(N), as the coordinates of N point scatters in the natural (drydock) ship coordinate system. $x_o$ is the long axis, $y_o$ is cross-ship, and $z_o$ is height.

Define the two degrees of rotation with respect to the radar as:

$$Aspect\ Angle = \phi(t)$$

$$Tilt\ Angle = \theta(t)$$

These two rotations convert [x₀, y₀, z₀] into the relative time-dependent coordinates [x, y, z].

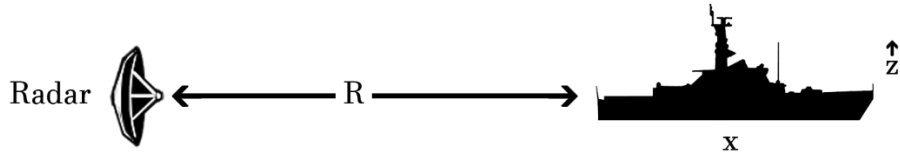

The transformation from [x₀, y₀, z₀] to the time-dependent positions [x, y, z] is:

$$\begin{pmatrix} x \\ y \\ z \end{pmatrix} = (Tilt * Aspect) \begin{pmatrix} x_o \\ y_o \\ z_o \end{pmatrix}$$

$$Tilt = \begin{pmatrix} \cos\theta & 0 & -\sin\theta \\ 0 & 1 & 0 \\ \sin\theta & 0 & \cos\theta \end{pmatrix}$$

$$Aspect = \begin{pmatrix} \cos\theta & -\sin\theta & 0 \\ \sin\theta & \cos\theta & 0 \\ 0 & 0 & 1 \end{pmatrix}$$

$\theta$ and $\phi$ are measured from the x-axis so x is the local range offset:

$$r = x_o \cos\theta \cos\phi - y_o \cos\theta \sin\phi - z_o \sin\theta$$

The first and second derivatives of $r$ with respect to time are:

$$\dot{r} = x_o[-\sin\theta \cos\phi\ \dot\theta - \cos\theta \sin\phi\ \dot\phi] \\ + y_o[\sin\theta \sin\phi\ \dot\theta - \cos\theta \cos\phi\ \dot\phi] \\ - z_o \cos\theta\ \dot\theta$$

$$\ddot{r} = x_\circ [-\cos\theta\cos\phi\,\dot{\theta}^2 + \sin\theta\sin\phi\,\dot{\theta}\dot{\phi}$$
$$-\sin\theta\cos\phi\,\ddot{\theta} + \sin\theta\sin\phi\,\dot{\theta}\dot{\phi}$$
$$-\cos\theta\cos\phi\,\dot{\phi}^2 - \cos\theta\sin\phi\,\ddot{\phi}]$$
$$+ y_\circ [\cos\theta\sin\phi\,\dot{\theta}^2 + \sin\theta\cos\phi\,\dot{\theta}\dot{\phi}$$
$$+ \sin\theta\sin\phi\,\ddot{\theta} + \sin\theta\cos\phi\,\dot{\theta}\dot{\phi}$$
$$+ \cos\theta\sin\phi\,\dot{\phi}^2 - \cos\theta\cos\phi\,\ddot{\phi}]$$
$$+ z_\circ [\sin\theta\,\dot{\theta}^2 - \cos\theta\,\ddot{\theta}]$$

Define the matrix inverse:

$$\begin{pmatrix} x_\circ \\ y_\circ \\ z_\circ \end{pmatrix} = M^{-1} \begin{pmatrix} r \\ \dot{r} \\ \ddot{r} \end{pmatrix}$$

The coefficients of $M$ can be written as:

$r$ Coefficients:
- $\cos\theta\cos\phi$
- $\cos\theta\sin\phi$
- $-\sin\theta$

$\dot{r}$ Coefficients:
- $-\sin\theta\cos\phi\,\dot{\theta} - \cos\theta\sin\phi\,\dot{\phi}$
- $\sin\theta\sin\phi\,\dot{\theta} - \cos\theta\cos\phi\,\dot{\phi}$
- $\cos\theta\,\dot{\theta}$

$\ddot{r}$ Coefficients:
- $[-(\dot{\theta}^2 + \dot{\phi}^2)\cos\theta\cos\phi + 2\sin\theta\sin\phi\,\dot{\theta}\dot{\phi} - \sin\theta\cos\phi\,\ddot{\theta} - \cos\theta\sin\phi\,\ddot{\phi}]$
- $[(\dot{\theta}^2 + \dot{\phi}^2)\cos\theta\sin\phi + 2\sin\theta\cos\phi\,\dot{\theta}\dot{\phi} + \sin\theta\sin\phi\,\ddot{\theta} - \cos\theta\cos\phi\,\ddot{\phi}]$
- $[\sin\theta\,\dot{\theta}^2 - \cos\theta\,\ddot{\theta}]$

These equations are used in multiple ways by the 3-D code. First, they are used in the built-in simulator to generate values of range, range rate, and acceleration from input time histories of tilt and aspect angle. Second, the inverse of the motion matrix is used to estimate the 3-D XYZ positions of the scatterers in the ship's natural coordinate system. Finally, using assumed error variances on range, Doppler, and acceleration the matrix equations are used to estimate the errors in the XYZ positions.